\documentclass[pra,twocolumn,numerical,floatfix]{revtex4-1}

\newcommand{\ket}[1]{| #1 \rangle}
\newcommand{\bra}[1]{\langle #1 |}

\usepackage[english]{babel}
\usepackage{amsmath,amssymb}
\usepackage{acronym}
\usepackage{setspace}
\usepackage{graphicx}

\usepackage{overpic}
\usepackage{rotating}
\usepackage{subfigure}

\usepackage{multirow}
\usepackage[colorlinks=true,bookmarks=true,citecolor=blue,urlcolor=blue]{hyperref} 

\begin{document}

\title{Quantum networking of microwave photons using optical fibers}

\author{B.D. Clader}
\email{dave.clader@jhuapl.edu}
\affiliation{The Johns Hopkins University Applied Physics Laboratory \\ 11100 Johns Hopkins Rd, Laurel, MD 20723, USA}

\begin{abstract}
We describe an adiabatic state transfer mechanism that allows for high-fidelity transfer of a microwave quantum state from one cavity to another through an optical fiber.  The conversion from microwave frequency to optical frequency is enabled by an optomechanical transducer.  The transfer process utilizes a combined dark state of the mechanical oscillator and fiber modes, making it robust against both mechanical and fiber loss.  We anticipate this scheme being an enabling component of a hybrid quantum computing architecture consisting of superconducting qubits with optical interconnects.
\end{abstract}

\maketitle

\acrodef{STIRAP}{stimulated Raman adiabatic passage}

%
%
%
%
\section{Introduction}
The rapidly expanding field of cavity optomechanics studies the interaction between mechanical and optical degrees of freedom in an optical resonator. This coupling occurs as radiation pressure from the confined cavity light acts to move the cavity mirrors, causing a shift in the cavity resonance. The first experiments to observe this interaction were done as far back as 1970 by Braginsky and coworkers \cite{Braginskii1970}. In this early work, the mirrors were relatively large suspended macroscopic objects. As early as 1996 it was suggested that similar systems could be fabricated on chip \cite{apl/69/18/10.1063/1.117548}.  More recent advances in nano fabrication culminated in the first ever observation of quantized mechanical motion in 2010 \cite{o2010quantum} (for recent reviews see Refs. \cite{Kippenberg:07, Kippenberg29082008, Physics.2.40, Aspelmeyer:10, 1303.0733}). 

When the mechanical oscillator's resonant frequency is much larger than the intrinsic loss rate of the cavity, sidebands in the optical cavity appear at the mechanical resonant frequency. In this so--called resolved sideband limit, one can perform sideband cooling of the mechanical oscillator by driving the cavity with a cooling laser tuned to the blue sideband \cite{PhysRevLett.99.093902, schliesser2008resolved}. As an additional constraint, when the optical cavity is strongly coupled to the mechanical resonator, such that the coupling rate is much greater than the intrinsic loss rates of the cavity and oscillator, quantum states can be exchanged between the cavity and mechanical modes \cite{PhysRevA.68.013808, verhagen2012quantum}.   Because the radiation pressure is so broadband, this ability to have strongly coupled systems can occur over a very large wavelength range. This fact has given rise to demonstrations of strong coupling and resolved sideband cooling in both the optical \cite{groblacher2009observation, schliesser2008resolved, PhysRevA.83.063835, chan2011laser, verhagen2012quantum} and microwave \cite{PhysRevLett.101.197203, o2010quantum, teufel2011sideband} frequency ranges.

This ability to strongly couple to both microwave and optical frequency cavities led to a proposal to use an optomechanical resonator as a wavelength convertor between quantum states of microwave and optical photons \cite{PhysRevA.82.053806}. Here the authors suggested state--swapping by using a sequence of $\pi/2$ optomechanical pulses, analogous to how one would perform state--swapping in an atomic system. Later proposals suggested that one could convert between microwave and optical frequencies by using an adiabatic dark state transfer scheme \cite{PhysRevLett.108.153603, PhysRevLett.108.153604}. While not converting between microwave and optical light, a recent experiment has demonstrated the use of a mechanical dark state to transfer quantum states between two optical cavities \cite{Dong21122012}. Alternative proposals for state swapping make use of the ability to entangle microwave and optical modes through a common optomechanical interface \cite{PhysRevA.84.042342, PhysRevLett.110.233602}, enabling continuous variable teleportation protocols to be utilized for state transfer \cite{PhysRevLett.109.130503}. Recent experimental results have realized broadband wavelength conversion, showing coherent wavelength conversion between microwave and optical frequencies \cite{hill2012coherent, bochmann2013nanomechanical}, even with room temperature devices \cite{bagci2013optical}.

A superconducting qubit's transition frequency is in the microwave regime \cite{Devoret08032013}, and recent advances in circuit quantum electrodynamics has created strongly coupled superconducting qubit and microwave resonators \cite{wallraff2004strong}. The ability to coherently convert between microwave and optical frequencies should enable one to develop a distributed quantum system using superconducting qubits and optical fiber connections. In doing so, one would improve the scalability of superconducting qubits in that the size constraints of a cryogenic system would no longer place a hard constraint on the number of qubits that one could couple. This idea to optically connect computational nodes of a quantum network is not new \cite{PhysRevLett.79.5242, PhysRevA.59.2659, PhysRevLett.96.010503}. It has been touted as a scalable architecture for ion--trap based quantum computers \cite{Monroe08032013}. The inability to reliably convert between microwave and optical photons has prevented such architectures for superconducting qubit systems. Recent proposals have appeared that analyzed the networking of solid-state qubits through optical fibers using optomechanical transducers \cite{PhysRevLett.105.220501, PhysRevA.84.042341}.

In this paper we present results that show that an adiabatic dark state transfer scheme, similar to the one outlined in Refs. \cite{PhysRevLett.108.153603, PhysRevLett.108.153604}, generalizes to fiber coupled systems. We show that this could enable high--fidelity state transfer between microwave cavities through optical fibers enabling a route towards optical networking of superconducting qubits. The frequency conversion on each side of the transfer is done using an optomechanical resonator, simultaneously coupled to both optical and microwave cavities. The optical cavities are then coupled via an optical fiber. A system schematic is shown in Fig. \ref{fig:Schematic}. Unlike recent proposals in Refs. \cite{PhysRevLett.105.220501, PhysRevA.84.042341}, the scheme outlined here considers a single mode fiber and the coupling of microwave photons to the optomechanical transducer (OMT) rather than the direct coupling of a qubit to the OMT. The use of a single mode fiber eliminates experimental complications that can prevent transfer of quantum states through the continuum \cite{PhysRevA.50.R913} except in very simple cases \cite{PhysRevLett.68.3523, *PhysRevA.47.571}.

Our scheme is similar to stimulated Raman adiabatic passage (STIRAP) \cite{PhysRevA.51.1578, RevModPhys.77.633}, used in quantum optics to perform state transfer between ground states of atoms without populating the excited state. As in the atomic case, our use of a dark state to perform the adiabatic passage reduces the effects of loss on the state transfer fidelity. We show, using analytical and numerical solutions, how one could perform such a transfer, and we calculate the transfer fidelities for a variety of input states. We predict that high-fidelity transfer is possible when 1) both the optical and microwave cavities are strongly coupled to the mechanical resonator, and 2) there is strong coupling between the optical cavities and the fiber.  

%
%
%
%
\section{Model}
\begin{figure}[h!]
\begin{center}
\begin{overpic}[width = 8.4cm]{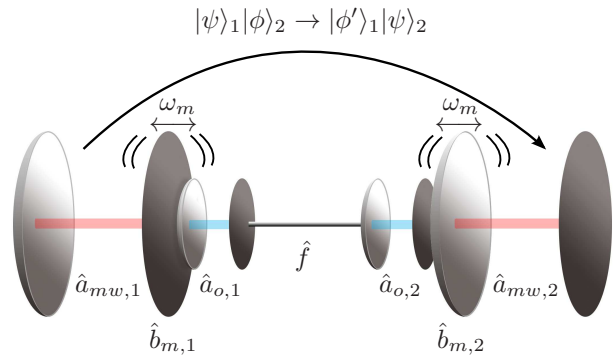}\end{overpic}
\put(-165,120){$\ket{\psi}_1\ket{\phi}_2 \to \ket{\phi^\prime}_1\ket{\psi}_2$}
\put(-210,20){$\hat{a}_{mw,1}$}
\put(-182,-1){$\hat{b}_{m,1}$}
\put(-163,20){$\hat{a}_{o,1}$}
\put(-127,30){$\hat{f}$}
\put(-95,20){$\hat{a}_{o,2}$}
\put(-73,-1){$\hat{b}_{m,2}$}
\put(-52,20){$\hat{a}_{mw,2}$}
\put(-182,81){$\longleftrightarrow$}
\put(-178,89){$\omega_m$}
\put(-74,81){$\longleftrightarrow$}
\put(-71,89){$\omega_m$}
\caption{\label{fig:Schematic}(Color Online) Schematic of the system being modeled. We consider two coupled cavity optomechanical systems denoted with labels 1 and 2. Each system consists of two cavities, one with a microwave resonant frequency denoted with red and cavity operator $\hat{a}_{mw,j}$, and one with an optical frequency denoted with blue and cavity operator $\hat{a}_{o,j}$. The cavities are couple via a common mechanical oscillator with mechanical frequency $\omega_m$ and operator $\hat{b}_{m,j}$. This system is then coupled to a second identical cavity optomechanical system via an optical fiber, with operator $\hat{f}$, that is connected with each optical cavity. This setup allows one to adiabatically transfer a quantum state stored in microwave cavity 1, denoted as $\ket{\psi}_1$, to microwave cavity 2 through the mechanical oscillators, optical cavities, and fiber. The adiabatic passage uses a dark state with minimal excitation in the mechanical and fiber modes, making it robust against loss in those modes.}
\end{center}
\end{figure}

Our model consists of two identical optomechanical sub-systems connected via an optical fiber. Each sub-system consists of a microwave cavity and an optical cavity connected via an optomechanical resonator as shown in Fig. \ref{fig:Schematic}. Such optomechanical devices have been demonstrated experimentally and shown to be capable of coherent conversion of microwave photons to optical photons \cite{hill2012coherent, bochmann2013nanomechanical}.

We model such a system with the following time-dependent Hamiltonian:
\begin{equation}
\label{eq:hamiltonian}
\hat{H} = \hat{H}_{om,1} + \hat{H}_{om,2} + \hat{H}_f + \hat{H}_{d}
\end{equation}
consisting of Hamiltonians for the optomechanical sub-systems 1 and 2 labelled $\hat{H}_{om,1}$ and $\hat{H}_{om,2}$ respectively, the fiber coupling term $\hat{H}_f$, and dissipation terms contained in $\hat{H}_{d}$ that couple the cavity, mechanical, and fiber modes to a continuum.
The linearized optomechanical sub-system Hamiltonian written in the interaction picture (see Appendix \ref{sec:h_derivation} for a detailed derivation) is
\begin{equation}
\label{eq:h_om}
H_{om,j} =  \hbar \sum_{i=o,mw} g_{i,j}(\hat{a}_{i,j}^\dagger \hat{b}_{m,j} + \hat{b}_{m,j}^\dagger \hat{a}_{i,j}),
\end{equation}
where the subscript $i=\{o,mw\}$ denotes the optical and microwave cavity and the subscript $j=\{1,2\}$ denotes the optomechanical sub-system. The annihilation operator of the cavity mode is $\hat{a}_{i,j}$, the annihilation operator of the mechanical mode is $\hat{b}_{m,j}$, and $g_{i,j}$ are the time-dependent optomechnanical coupling parameters (see Appendix \ref{sec:h_derivation} for an explanation of the how to tune $g$ parameters by varying a classical cavity driving field).

The optical cavity to fiber coupling Hamiltonian is \cite{PhysRevLett.79.5242}
\begin{align}
\label{eq:h_fiber}
\hat{H}_f & = \hbar \sum_{k}\Delta_k\hat{f}_k^\dagger\hat{f}_k \\ \nonumber
& + \left\{g_{f}\sum_{k}[\hat{a}_{o,1} + (-1)^k \hat{a}_{o,2}]\hat{f}_k^\dagger + \textnormal{H.c}\right\},
\end{align}
where $\Delta_k = \omega_k - \omega_{oc}$ is the frequency difference between the $k^{th}$ fiber mode and the cavity mode, $g_{f}$ is the coupling strength between the fiber and the optical cavities, and $\hat{f}_k$ is the annihilation operator of the $k^{th}$ fiber mode.  

The finite length of the fiber implies a quantization of the modes of the fiber with frequency spacing $2\pi c/l$ where $l$ is the length of the fiber. Therefore the number of modes that significantly interact with the cavity is on the order of $N = \Gamma l/ 2\pi c$, where $\Gamma$ is the decay rate of the cavity fields into a continuum of fiber modes \cite{PhysRevLett.79.5242}.  For short enough fibers $N \lesssim 1$, and one can employ the short--fiber limit \cite{PhysRevLett.79.5242, PhysRevA.59.2659, PhysRevLett.96.010503} and only keep the single fiber mode that is resonant with the optical cavity in Eq. \eqref{eq:h_fiber}. In this limit the fiber Hamiltonian becomes
\begin{equation}
\label{eq:h_fiber_short}
\hat{H}_f =\hbar g_f\left[(\hat{a}_{o,1} - \hat{a}_{o,2})\hat{f}^\dagger + (\hat{a}_{o,1}^\dagger - \hat{a}_{o,2}^\dagger)\hat{f}\right].
\end{equation}

We model dissipation by allowing the optical and microwave cavities, fiber, and mechanical modes to each couple to independent continua of harmonic oscillator modes.  The dissipation Hamiltonian is then composed of $\hat{H}_d = \sum_{i=1}^{2}(\hat{H}_{do,i} + \hat{H}_{dm,i} + \hat{H}_{dmw,i}) + \hat{H}_{df},$ where $\hat{H}_{do,i}$ is the coupling of the $i^{th}$ optical cavity to the bath, $\hat{H}_{dm,i}$ the mechanical coupling, $\hat{H}_{dmw,i}$ the microwave cavity coupling, and finally $\hat{H}_{df}$ describes the coupling of the fiber modes to its dissipation channel.  These Hamiltonians are all similar in form. For simplicity we write just the one for the optical cavity. Within the rotating wave approximation this Hamiltonian is

\begin{align}
\label{eq:h_dissipation}
\hat{H}_{do,i} & = \hbar\int_{-\infty}^{\infty}d\omega (\omega - \omega_{o,i})\hat{c}_{o,i}^\dagger(\omega)\hat{c}_{o,i}(\omega) \\ \nonumber
& + i\hbar\int_{-\infty}^{\infty}d\omega g_{do,i}(\hat{a}_{o,i}^\dagger\hat{c}_{o,i} - \hat{a}_{o,i}\hat{c}_{o,i}^\dagger),
\end{align}

where $\omega_{o,i}$ corresponds to the resonant frequency of the optical cavity, $\hat{c}_{o,i}$ is the annihilation operator of the optical cavity bath, and the coupling coefficients is $g_{do,i}$.  We assume the canonical commutation relations for the bath operators $[\hat{c}_{o,i}(\omega),\hat{c}_{o,j}(\omega^\prime)^\dagger] = \delta_{i,j}\delta(\omega - \omega^\prime)$. A similar Hamiltonian exists for the microwave bath with bath operators $\hat{c}_{mw,i}$, as well as the mechanical and fiber modes with the same convention for those bath operators.

We now proceed to calculate the Heisenberg equations of motion for each of the various operators and eliminate the bath operators as done using the standard input--output formalism techniques \cite{gardiner2004quantum}.  The operator equation for the $i^{th}$ optical cavity bath is 
\begin{equation}
\label{eq:bath_equations}
i\frac{\partial\hat{c}_{o,i}}{\partial t} = (\omega - \omega_{o,i})\hat{c}_{o,i} - i g_{do,i}\hat{a}_{o,i}.
\end{equation}
Equations identical in form to Eq. \eqref{eq:bath_equations} can be similarly obtained for the microwave cavities, mechanical oscillators, and fiber mode baths.  Meanwhile, the optical and microwave cavities, mechanical oscillator, and fiber mode equations of motion are
\begin{subequations}
\label{eq:operator_equations}
\begin{align}
i\frac{\partial\hat{a}_{o,i}}{\partial t} & = g_{o,i}\hat{b}_{m,i} \pm g_f\hat{f} + i \int_{-\infty}^{\infty}d\omega g_{do,i}\hat{c}_{o,i} \label{eq:optical_cavity} \\
i\frac{\partial\hat{b}_{m,i}}{\partial t} & = g_{o,i}\hat{a}_{o,i} + g_{mw}\hat{a}_{mw,i} + i \int_{-\infty}^{\infty}d\omega g_{dm,i}\hat{c}_{m,i} \\
i\frac{\partial\hat{a}_{mw,i}}{\partial t} & = g_{mw,i}\hat{b}_{mw,i} + i \int_{-\infty}^{\infty}d\omega g_{dmw,i}\hat{c}_{mw,i} \\
i\frac{\partial\hat{f}}{\partial t} & = g_{f}\hat{a}_{o,1} -  g_{f}\hat{a}_{o,2} + i \int_{-\infty}^{\infty}d\omega g_{df}\hat{c}_{f},
\end{align}
\end{subequations}
where the $\pm$ in Eq. \eqref{eq:optical_cavity} is for cavity 1 and 2 respectively.

We integrate Eq. \eqref{eq:bath_equations} directly giving
\begin{align}
\label{eq:bath_int}
\hat{c}_{o,i}(t) &= e^{-i(\omega - \omega_{o,i})(t-t_0)}\hat{c}_{o,i}(t_0) \\ \nonumber
& - g_{do,i}\int_{t_0}^{t}dt^\prime e^{-i(\omega-\omega_{o,i})(t-t^\prime)}\hat{a}_{o,i}(t^\prime),
\end{align}
and insert this solution along with the similar forms for the other terms into Eqs. \eqref{eq:operator_equations}.  We assume a Markovian bath such that the various coupling terms $g_{do,j}$ are frequency independent, yielding the following set of quantum Langevin equations:
\begin{equation}
\label{eq:operator_equations3}
i\frac{\partial\vec{v}(t)}{\partial t} = M(t)\vec{v}(t) + i\sqrt{K}\vec{v}_{in}(t),
\end{equation}
where we define $$\vec{v} = (\hat{a}_{o,1},\hat{b}_{m,1},\hat{a}_{mw,1},\hat{f},\hat{a}_{o,2},\hat{b}_{m,2},\hat{a}_{mw,2})^T,$$ $$\vec{v}_{in} = (\hat{a}_{o,1,in},\hat{b}_{m,1,in},\hat{a}_{mw,1,in},\hat{f}_{in},\hat{a}_{o,2,in},\hat{b}_{m,2,in},\hat{a}_{mw,2,in})^T,$$ $$K = \textnormal{diag}(\kappa_o,\kappa_m,\kappa_{mw},\kappa_f,\kappa_o,\kappa_m,\kappa_{mw})^T/2\pi,$$ and the dynamics matrix $M$ is
\begin{widetext}
\begin{equation}
\label{eq:m_matrix}
M(t) = \left(\begin{array}{ccccccc}
-i \kappa_o/2 & g_{o,1}(t) & 0 & g_{f} & 0 & 0 & 0 \\
g_{o,1}(t) & -i \kappa_m/2 & g_{mw,1}(t) & 0 & 0 & 0 & 0 \\
0 & g_{mw,1}(t) & -i \kappa_{mw}/2 & 0 & 0 & 0 & 0 \\
g_f & 0 & 0 & -i \kappa_f/2 & -g_f & 0 & 0 \\
0 & 0 & 0 & -g_f & -i\kappa_o/2 & g_{o,2}(t) & 0 \\
0 & 0 & 0 & 0 & g_{o,2}(t) & -i\kappa_m/2 & g_{mw,2}(t) \\
0 & 0 & 0 & 0 & 0 & g_{mw,2}(t) & -i\kappa_{mw} \end{array}\right).
\end{equation}
\end{widetext}
The noise terms $$\hat{a}_{j,i,in} = \int_{-\infty}^{\infty}d\omega e^{-i(\omega-\omega_j)(t-t_0)}\hat{c}_{j,i}(t_0)$$ describe the bath fluctuations for the cavities with $j=\{o,mw\}$.  Similar definitions apply for the mechanical and fiber modes.  We assume that each optical mode's  noise distribution is white and Markovian such that $$\langle \hat{a}_{j,i,in}(t^\prime)^\dagger \hat{a}_{m,n,in}(t)\rangle = \delta_{j,m}\delta_{i,n}\delta(t-t^\prime).$$  We assume thermal noise for the mechanical oscillators with $$\langle \hat{b}_{m,j,in}(t^\prime)\hat{b}_{m,k,in}^\dagger(t)\rangle = (N_{th}+1)\delta_{j,k}\delta(t-t^\prime),$$ where $N_{th}$ is the thermal phonon occupation number of the bath.  In addition we define the cavity decay rates as $g_{do,i}^2 \equiv \kappa_o/2\pi$ with similar definitions for the microwave cavity, mechanical oscillator, and fiber modes. We note here that we have removed the subsystem label from the decay rates as we are assuming identical optomechanical devices on each end of the fiber.

Eqs. \eqref{eq:operator_equations3} and \eqref{eq:m_matrix} are our starting point for the next section where we demonstrate how one can achieve high-fidelity state transfer between the microwave cavity modes by adiabatically varying the coupling rates.  We have explicitly included the time--dependence of the various optomechanical coupling rates as these will be the knobs we use to engineer such a state transfer protocol.  In an experimental setting, these coupling rates are straightforward to tune, as they are related to the power of the optomechanical driving laser (see Appendix \ref{sec:h_derivation} for more details).

%
%
%
%
\section{Quantum State Transfer Protocol}\label{sec:state_transfer}
We now proceed to describe an adiabatic dark state transfer protocol that enables one to perform the following state transformation:
\begin{equation}
\label{eq:state_transfer}
\ket{\psi}_1\ket{\phi}_2 \to \ket{\phi^\prime}_1\ket{\psi}_2,
\end{equation}
where the subscripts 1 and 2 refer to microwave cavity 1 and 2 respectively. Eq. \eqref{eq:state_transfer} implies we intend to transfer the arbitrary quantum microwave cavity state from cavity 1 to cavity 2. This is not a swap gate in that the final state of microwave cavity 1 is not the initial state of cavity 1.  Rather, it consists of a transformation of the initial states of the various other optical cavity, fiber, and mechanical oscillator initial states.

To see how such a transfer is possible, we rewrite Eq. \eqref{eq:operator_equations3} as
\begin{equation}
\label{eq:diagonal_Langevin}
i \frac{\partial \vec{\tilde{v}}(t)}{\partial t} = i \frac{\partial U(t)^\dagger}{\partial t}U\vec{\tilde{v}}(t) + \Lambda(t)\vec{\tilde{v}}(t) + i\sqrt{K}\vec{\tilde{v}}_{in}(t),
\end{equation}
where $\Lambda(t) = U(t)^\dagger M(t)U(t)$ is the diagonalized version of $M(t)$, $U(t)$ contains the normalized eigenvectors of $M(t)$ in each column, and $\vec{\tilde{v}}(t) = U(t)^\dagger \vec{v}(t)$.  The adiabatic condition implies that the time dependent couplings in Eq. \eqref{eq:m_matrix} obey the inequality $|\partial g_{i,j}/\partial t| \ll g_{0,j}^2$ where $g_{0,j}(t)^2 = g_{o,j}(t)^2 + g_{mw,j}(t)^2$.  When this inequality holds, we can neglect the first term on the right hand side of Eq. \eqref{eq:diagonal_Langevin} giving
\begin{equation}
\label{eq:diagonal_Langevin2}
i \frac{\partial \vec{\tilde{v}}(t)}{\partial t} \approx \Lambda(t)\vec{\tilde{v}}(t) + i\sqrt{K}\vec{\tilde{v}}_{in}(t),
\end{equation}
the solution of which is given by
\begin{equation}
\label{eq:solved_Langevin}
\vec{\tilde{v}}(t) = e^{-i\int_{0}^{t}dt^\prime \Lambda(t^\prime)}\vec{\tilde{v}}(0) + i \sqrt{K} \int_{0}^{t}dt^\prime e^{-i\int_{t^\prime}^{t}dt^{\prime\prime}\Lambda(t^{\prime\prime})}\vec{\tilde{v}}_{in}(t^\prime).
\end{equation}

To integrate Eq. \eqref{eq:solved_Langevin} requires us to solve for the eigenvectors and eigenvalues of the matrix in Eq. \eqref{eq:m_matrix}. Unfortunately this is difficult in general.  We can gain insight into the dynamics by setting each of the cavity--bath loss terms on the diagonal to 0.  In that case, solving for the full eigensystem is still difficult, as it contains a cubic equation.  There is one solution that can be easily obtained.  It is straightforward to show that the vector
\begin{equation}
\label{eq:dark_state}
\psi_d(t) = \frac{1}{g_T(t)^2}\left(\begin{array}{c}
-g_{mw,1}(t)g_{mw,2}(t) \\
0 \\
g_{o,1}(t)g_{mw,2}(t) \\
0 \\
-g_{mw,1}(t)g_{mw,2}(t) \\
0 \\
g_{mw,1}(t)g_{o,2}(t)\end{array}\right)
\end{equation}
is an eigenvector of the matrix $M(t)$ (with all diagonal terms set to 0) with eigenvalue 0, where $g_T(t)^4 = g_{mw,1}(t)^2g_{o,2}(t)^2 + 2g_{mw,1}(t)^2g_{mw,2}(t)^2 + g_{o,1}(t)^2g_{mw,2}(t)^2$.  In other words, $\psi_d(t)$ is a dark state of the lossless system.  To see how loss affects this dark state, we use Eq. \eqref{eq:dark_state} along with the diagonal terms in Eq. \eqref{eq:m_matrix} to calculate the dark state eigenvalue to first order in perturbation theory giving\begin{align}
\label{eq:dark_state_ev}
\lambda_d &= -\frac{i}{2g_T(t)^4} \big[2\kappa_{o} g_{mw,1}(t)^2g_{mw,2}(t)^2 \\ \nonumber 
&+ \kappa_{mw}g_{o,1}(t)^2g_{mw,2}(t)^2 + \kappa_{mw} g_{mw,1}(t)^2g_{o,2}(t)^2\big].
\end{align}
One can see that only the optical and microwave decay rates appear in the dark state eigenvalue. The non-zero value indicates that loss causes coupling of the dark state to other modes that are, in turn, coupled to the environment, resulting in imperfect state transfer fidelity.

Insight into the adiabatic transfer scheme can be seen by examining Eq. \eqref{eq:dark_state}.  One uses a ``counter-intuitive'' pulse sequence that begins at $t=0$ as $$g_{mw,1}(0) = g_{o,2}(0) = 0 \quad\textnormal{and}\quad g_{o,1}(0) = g_{mw,2}(0) = g$$ and ends at $t=T$ with $$g_{mw,1}(T) = g_{o,2}(T) = g \quad\textnormal{and}\quad g_{o,1}(T) = g_{mw,2}(T) = 0.$$  With such a sequence it is straightforward to show that $\psi_d(0) = (0,0,1,0,0,0,0)^T$ and $\psi_d(T) = (0,0,0,0,0,0,1)^T$.  This implies that
\begin{equation}
\label{eq:U_0}
U(0)^\dagger = \left(\begin{array}{ccccccc}
0 & 0 & 1 & 0 & 0 & 0 & 0 \\
* & * & * & * & * & * & * \\
* & * & * & * & * & * & * \\
* & * & * & * & * & * & * \\
* & * & * & * & * & * & * \\
* & * & * & * & * & * & * \\
* & * & * & * & * & * & * \end{array}\right)
\end{equation}
and
\begin{equation}
\label{eq:U_T}
U(T)^\dagger = \left(\begin{array}{ccccccc}
0 & 0 & 0 & 0 & 0 & 0 & 1 \\
* & * & * & * & * & * & * \\
* & * & * & * & * & * & * \\
* & * & * & * & * & * & * \\
* & * & * & * & * & * & * \\
* & * & * & * & * & * & * \\
* & * & * & * & * & * & * \end{array}\right),
\end{equation}
where the $*$ terms indicate the other unknown eigenstates of $M(t)$. At time $t=0$ the dark mode is the first microwave cavity mode since $\tilde{v}_1(0) = [U(0)^\dagger \vec{v}(0)]_1 = \hat{a}_{mw,1}$ and at time $t=T$ the dark mode is the second microwave cavity mode since $\tilde{v}_1(T) = [U(T)^\dagger \vec{v}(T)]_1 = \hat{a}_{mw,2}$.  Therefore, as long as one performs this passage adiabatically, which in this case implies $T \gg 1/g_T(t)$, the system will remain in the dark state while the excitation from the first microwave cavity will be transferred to the second microwave cavity. This scheme requires fully coherent microwave and optical cavity driving sources. While experimentally challenging, recent advances in frequency combs suggest that this does not present an insurmountable technical hurdle \cite{[{See e.g., }] Diddams:10, *Kippenberg29042011}.

%
%
%
%
\section{State Transfer Fidelity Model}
To see how well this transfer protocol works, we wish to estimate the state transfer fidelity for a variety of input quantum states, including squeezed--coherent states and qubit states.  We define squeezed--coherent states in the usual way: $\ket{\alpha,s} = \hat{D}(\alpha)\hat{S}(r)\ket{0}$ with $D(\alpha) = \exp(\alpha \hat{a}_{mw,1}^\dagger - \alpha^*\hat{a}_{mw,1})$ the displacement operator and $\hat{S}(r) = \exp(\frac{1}{2}r^*\hat{a}_{mw,1}^2 + \frac{1}{2}r(\hat{a}_{mw,1}^\dagger)^2)$  the squeezing operator \cite{loudon2000quantum}.  We define a qubit state as an arbitrary superposition of the $\ket{0}$, $\ket{1}$ Fock--states, $\alpha\ket{0} + \beta\ket{1}$.

Because the matrix $M(t)$ in Eq. \eqref{eq:m_matrix} is difficult to diagonalize in general, we obtain the state transfer fidelity by integrating the solution to Eq. \eqref{eq:operator_equations3} numerically. We outline our numerical solution method in Appendix \ref{sec:numerical_solution}. The numerical solution we integrate is given by
\begin{align}
\label{eq:solved_Langevin_general}
\vec{v}(t) = & e^{-i\int_{0}^{t}dt^\prime M(t^\prime)}\vec{v}(0) \\ \nonumber
& + i \sqrt{K} \int_{0}^{t}dt^\prime e^{-i\int_{t^\prime}^{t}dt^{\prime\prime}M(t^{\prime\prime})}\vec{v}_{in}(t^\prime).
\end{align}
We note that our numerical solution to Eq. \eqref{eq:solved_Langevin_general} is exact in the sense that we do not make any adiabatic approximations. This allows us to see what affect non--adiabaticy has on the state--transfer fidelity.

The fidelity of two mixed quantum states given by density matrices $\rho_1$ and $\rho_2$ is given by \cite{doi:10.1080/09500349414552171} 
\begin{equation}
\label{eq:fidelity}
F(\rho_1,\rho_2) = \textnormal{Tr}\left[\sqrt{\rho_1}\rho_2\sqrt{\rho_1}\right].
\end{equation}
Isar showed that, because Gaussian states, like the squeezed--coherent states above, are defined entirely by their covariance matrices, the fidelity between two Gaussian states can be written as \cite{isar2009}
\begin{equation}
\label{eq:gaussian_fidelity}
F_G = \frac{1}{\sqrt{\textnormal{det}[(A_1 + A_2)/2]}}e^{-\beta^T(A_1+A_2)^{-1}\beta}.
\end{equation}
We set our initial state $\rho_1 = \ket{\psi_1}\bra{\psi_1}$ to one of the pure quantum states defined above. After integrating Eq. \eqref{eq:solved_Langevin_general}, we extract the last component in $\vec{v}(t)$ that gives us the state in microwave cavity 2, since $[\vec{v}(t)]_7 = \hat{a}_{mw,2}$. The terms $A_j$ are the covariance matrices of the state quadratures $\hat{q}_j = (\hat{a}_j+\hat{a}_j^\dagger)/\sqrt{2}$ and $\hat{p}_j = -i(\hat{a}_j-\hat{a}_j^\dagger)/\sqrt{2}$ and are defined as
\begin{equation}
\label{eq:covariance_matrix}
A_j = \left(\begin{array}{cc}
2\sigma_{q_jq_j} & \sigma_{p_jq_j} + \sigma_{q_jp_j} \\
\sigma_{p_jq_j} + \sigma_{q_jp_j} & 2\sigma_{q_jq_j}
\end{array}\right),
\end{equation}
where $\sigma_{q_jq_j} = \langle \hat{q}_j\hat{q}_j\rangle - \langle \hat{q}_j\rangle^2$ and similarly for $\hat{p}_j$, while the covariance is $\sigma_{p_jq_j} = \langle \hat{p}_j\hat{q}_j\rangle - \langle \hat{p}_j\rangle \langle \hat{q}_j\rangle$. The term $\beta = \alpha_2 - \alpha_1$ is the difference of the mean value of the quadrature amplitudes with
\begin{equation}
\label{eq:alpha}
\alpha_i = \left(\begin{array}{c}
\langle \hat{q}_i \rangle \\
\langle \hat{p}_i \rangle
\end{array}\right).
\end{equation}

For qubit states we calculate the state-transfer fidelity using the formalism of Julsgaard and M\o lmer \cite{PhysRevA.89.012333}.  They derived the the fidelity between two qubit states as a rather complicated expression given by
\begin{widetext}
\begin{align}
\label{eq:qubit_fidelity}
F_q = & \frac{1}{6\sqrt{(\sigma_1^2 + \frac{1}{2})(\sigma_2^2 + \frac{1}{2})}}\bigg\{ 3 + \frac{3(\sigma_1^2\sigma_2^2 - \frac{1}{4})}{(\sigma_2^2 + \frac{1}{2})(\sigma_2^2 + \frac{1}{2})} + \frac{\textnormal{Re}[C+\tilde{D}^*]}{\sigma_1^2 + \frac{1}{2}} + \frac{\textnormal{Re}[C-\tilde{D}^*]}{\sigma_2^2 + \frac{1}{2}} \\ \nonumber
& -\frac{|C+\tilde{D}^*|^2(\sigma_1^2 - 1)}{(\sigma_1^2 +\frac{1}{2})^2} -\frac{|C-\tilde{D}^*|^2(\sigma_2^2 - 1)}{(\sigma_2^2 +\frac{1}{2})^2} - \frac{|C+\tilde{D}^*|^2(\sigma_2^2 - \frac{1}{2}) + |C-\tilde{D}^*|^2(\sigma_1^2 - \frac{1}{2})}{2(\sigma_1^2 +\frac{1}{2})(\sigma_2^2 +\frac{1}{2})}\bigg\}.
\end{align}
\end{widetext}
The various parameters are defined as follows:
\begin{align}
\label{eq:qubit_fidelity_params}
& \sigma_1^2 = \bar{\sigma}^2 + \delta\sigma^2 \\ \nonumber
& \sigma_2^2 = \bar{\sigma}^2 - \delta\sigma^2 \\ \nonumber
& \tan(2\theta) = \frac{\sigma_{q_2p_2} + \sigma_{p_2q_2}}{\sigma_{q_2q_2}^2 - \sigma_{p_2p_2}} \\ \nonumber
& \bar{\sigma}^2 = \frac{\sigma_{q_2q_2}^2 + \sigma_{p_2p_2}^2}{2} \\ \nonumber
& \delta\sigma^2 = \sqrt{\frac{1}{4}(\sigma_{q_2q_2} - \sigma_{p_2p_2})^2 + \frac{1}{4}(\sigma_{q_2p_2} + \sigma_{p_2q_2})^2}.
\end{align}
The $C$ and $\tilde{D}$ terms are given by
\begin{align}
\label{eq:c_d_defs}
C &= \frac{1}{2}(B_{11} - i B_{12} + i B_{21} + B_{22}) \\ \nonumber
\tilde{D} &= \frac{1}{2}(B_{11} + i B_{12} + i B_{21} - B_{22})e^{-2i\theta}.
\end{align}
The $B_{ij}$ terms come from the propagator $\exp(-i\int_0^t dt^\prime M(t^\prime))$ defined in \eqref{eq:solved_Langevin_general}.  We write the propagation equation analogous to Eq. \eqref{eq:solved_Langevin_general} for the $\hat{q}_{mw,2}$ and $\hat{p}_{mw,2}$ quadrature variables alone. This yields an equation of the form
\begin{equation}
\label{eq:xp_propagator}
\left( \begin{array}{c}
\hat{q}_{mw,2} \\
\hat{p}_{mw,2}
\end{array}\right) = 
\left( \begin{array}{cc}
B_{11} & B_{12} \\
B_{21} & B_{22} \end{array} \right)
\left( \begin{array}{c}
\hat{q}_{mw,1} \\
\hat{p}_{mw,2} \end{array}\right) + 
\left(\begin{array}{c}
\hat{F}_{q_{mw,2}} \\
\hat{F}_{p_{mw,2}} \end{array}\right),
\end{equation}
where $\hat{F}_{q_{mw,2}}$ and $\hat{F}_{p_{mw,2}}$ are the associated noise operators for the $\hat{q}_{mw,2}$ and $\hat{p}_{mw,2}$ quadrature variables respectively along with coupling to any other degrees of freedom besides the microwave cavity. These $B_{ij}$ terms are used in Eq. \eqref{eq:c_d_defs} to calculate the qubit transfer fidelity.

%
%
%
%
\section{State Transfer Fidelity Results}
With this framework in place, we proceed to calculate the state transfer fidelity by numerically integrating Eq. \eqref{eq:solved_Langevin_general}. As was shown in Sec. \ref{sec:state_transfer}, our transfer protocol requires that we use a counter--intuitive sequence optomechanical coupling terms.  One is free to choose the pulse shapes.  We choose the simplest scheme that satisfies the constraints implied by Eq. \eqref{eq:dark_state}.  Namely we take the pulses to be
\begin{align}
\label{eq:pulse_shapes}
- g_{o,1}(t) & = g_{mw,2}(t) = g\left(1 - \frac{t}{T}\right) \\ \nonumber
g_{mw,1}(t) & = -g_{o,2}(t) = g \frac{t}{T},
\end{align}
where $0 \le t \le T$ and $g$ is the maximum pulse strength. For all pulses we see that $|\partial g_{i,j}/\partial t|^2 = g^2/T^2$, therefore the adiabatic condition requires that $gT/2 \gg 1$.

\begin{figure}
\centering
\includegraphics[width=8.4cm]{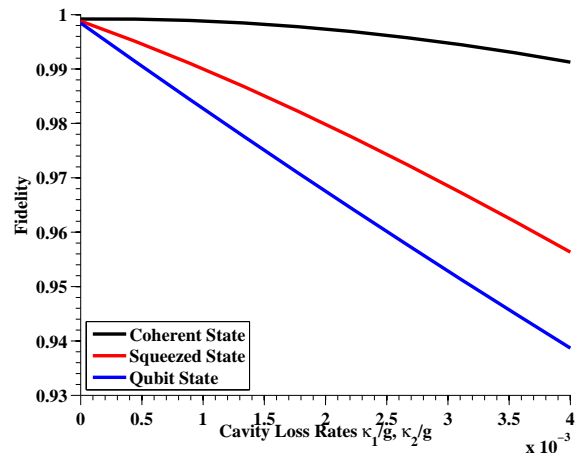}
\caption{(Color Online) Plot of the state transfer fidelity for a coherent state $\ket{\alpha,s} = \ket{1,0}$ (black - top), a squeezed coherent state $\ket{\alpha,s} = \ket{1,0.4}$ (red - middle), and a qubit state $\alpha\ket{0}+\beta\ket{1}$ (blue - bottom). Both optical and microwave cavity loss rates are varied equally as specified by the $x$ axis.  We set the fiber and optomechanical loss rates to $\kappa_f / g= \gamma_m / g= 0.002$, the total time to $g T/2 = 25$, and the thermal population to $N_{th} = 10$.}
\label{fig:cavity_loss}
\end{figure}

\begin{figure}
\centering
\includegraphics[width=8.4cm]{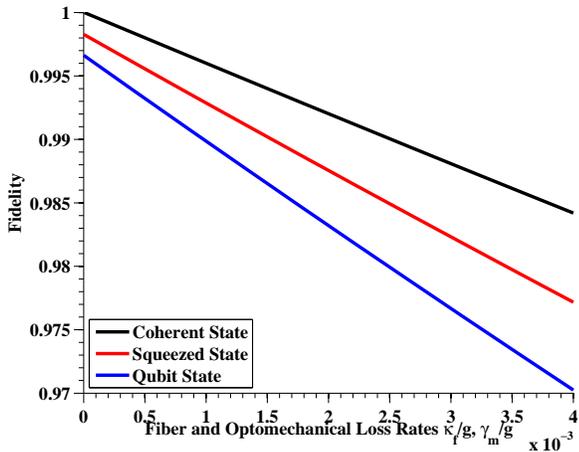}
\caption{(Color Online) Plot of the state transfer fidelity for a coherent state $\ket{\alpha,s} = \ket{1,0}$ (black - top), a squeezed coherent state $\ket{\alpha,s} = \ket{1,0.4}$ (red - middle), and a qubit state $\alpha\ket{0}+\beta\ket{1}$ (blue - bottom). Both fiber and mechanical oscillator loss rates are varied equally as specified by the $x$ axis.  We set the optical and microwave loss rates to $\kappa_o / g= \kappa_{mw} / g= 0.002$, the total time to $g T/2 = 25$, and the thermal population to $N_{th} = 10$.}
\label{fig:fiber_om_loss}
\end{figure}

We use the resulting solution to estimate the transfer fidelity for Gaussian states using the formula given in Eq.\eqref{eq:gaussian_fidelity} and for qubit states using the formula given in Eq. \eqref{eq:qubit_fidelity}. One would expect that the fidelity is more sensitive to changes in the cavity loss rates, compared to the mechanical and fiber loss rates, since we are using a dark state that is decoupled from the fiber and mechanical modes to lowest order. To test this, we fix the fiber loss and optomechanical loss terms, and look at how the state transfer fidelity scales as the microwave and cavity loss rates vary.  These results are plotted in Fig. \ref{fig:cavity_loss}. Next we fix the optical and microwave cavity loss rates and vary the fiber and mechanical loss rates.  These results are plotted in Fig. \ref{fig:fiber_om_loss}.  

As seen from the two figures, the results do indicate that the transfer fidelity is more sensitive to cavity loss than to fiber and mechanical loss. As the cavity loss rate approaches zero, the transfer fidelity is nearly perfect even though we have a non-zero mechanical and fiber loss rate in Fig. \ref{fig:cavity_loss}. In Fig. \ref{fig:fiber_om_loss} we see that with non-zero cavity loss rates, nearly perfect transfer fidelity is only possible for the coherent state input. Additionally the fidelity gets worse more quickly when increasing cavity loss rates than the corresponding mechanical and fiber loss rates. We also note that in both cases, the coherent state is least susceptible to loss, while the squeezed state and qubit state are progressively more prone to loss. This is expected as the coherent state a nearly classical state, while the squeezed state and qubit state are progressively more ``quantum'', causing them to be more fragile.

\begin{figure}
\centering
\includegraphics[width=8.4cm]{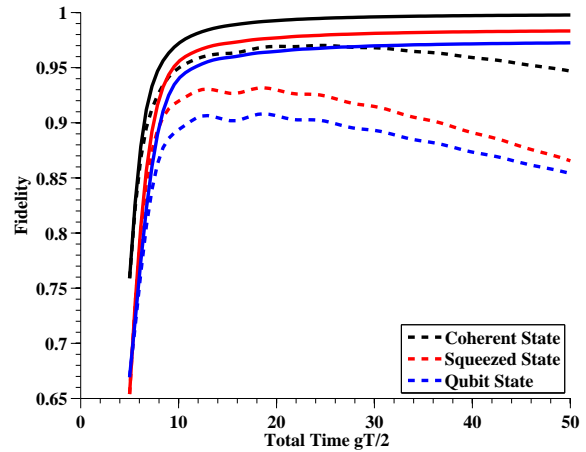}
\caption{(Color Online) Plot of the state transfer fidelity for a coherent state $\ket{\alpha,s} = \ket{1,0}$ (black - top), a squeezed coherent state $\ket{\alpha,s} = \ket{1,0.4}$ (red - middle), and a qubit state $\alpha\ket{0}+\beta\ket{1}$ (blue - bottom). Here we study the effects of adiabaticity by varying $g$ (solid curves) and $T$ (dashed curves).  We set the optical and microwave loss rates to $\kappa_o / g= \kappa_{mw} / g = \kappa_f /g = \gamma_m / g = 0.05$ for the dashed curves and $\kappa_o T= \kappa_{mw} T = \kappa_f T = \gamma_m T = 0.05$ for the solid curves. The thermal population is set to $N_{th} = 10$. As seen in the figure, only when $gT/2 \gtrsim 10$ can we achieve high--transfer fidelity state--transfer. This is expected due to requirements for adiabaticity. However, while it is always better to increase the optomechanical coupling rate $g$, one cannot increase $T$ indefinitely as eventually photon loss becomes an issue leading to an optimal transfer time.}
\label{fig:varying_adiabaticity}
\end{figure}

For the pulse shapes defined in Eq. \eqref{eq:pulse_shapes}, the adiabatic condition requires that $gT/2 \gg 1$. To satisfy this condition one can choose to increase either a large $g$ or $T$. We will show that it is always desirable to have $g$ large, while increasing $T$ enhances susceptibility to loss causing reduced fidelity. To see this, we vary $g$ and $T$ defined in Eq. \eqref{eq:pulse_shapes}. We plot the transfer fidelity in Fig. \ref{fig:varying_adiabaticity} for the same three input states as before. In the figure, we plot two sets of curves. The solid curves correspond to varying $g$ with fixed $T$, while the dashed lines correspond to varying $T$ with fixed $g$.

Clearly, for small $gT/2$, non--adiabaticity has a strongly negative influence on the transfer fidelity. Increasing $g$ always increases the fidelity up to the limit imposed by the loss rates. In principle one can always increase $g$ while decreasing $T$ to make loss negligible. In this limit the transfer fidelity approaches 1. Conversely, as $T$ is increased, the transfer fidelity initially goes up due to the satisfying of the adiabatic condition. As one continues to increase $T$ cavity loss begins to lower the fidelity. The longer the transfer sequence, the more likely photon loss will occur. For this reason, eventually the fidelity begins to go down as $T$ is continuously increased. This feature leads to an optimal transfer time $T$ that is governed by the coupling parameter and the various loss rates.

The parameters we have chosen here that lead to high transfer fidelity require one to have both the optical cavity and microwave cavity to be strongly coupled to the mechanical oscillator.  In addition, we require that the fiber be strongly coupled to the optical cavity.  We find that the ratio of the optomechanical coupling to cavity loss rate that produces high-fidelity transfers is around $g/\kappa \gtrsim 10$. These are challenging requirements for an experimental implementation, as the best recent experiments are closer to $g/\kappa \gtrsim 1$ for optical \cite{groblacher2009observation, verhagen2012quantum} and microwave setups \cite{o2010quantum, teufel2011sideband}. However, given the remarkable improvements over the last few years, we are optimistic that these requirements are plausible for a future experimental setup to meet.

%
%
%
%
\section{Conclusions}

We have derived the dynamical model that describes two microwave--mechanical--optical systems coupled through an optical fiber.  We derive solutions to this model that describe an adiabatic state transfer from one microwave cavity to the other through a quantum dark state. This state transfer is somewhat analogous to stimulated Raman adiabatic passage familiar to quantum optics.  Because the dark state is decoupled from the mechanical and fiber modes, the state transfer is relatively insensitive to loss in those modes.

Current superconducting qubit architectures make use of microwave cavities to interact with the qubits for state preparation, measurement, and control. The ability to transfer microwave quantum states through optical fibers will enable one to couple remote superconducting qubit systems. This will allow one to design a distributed superconducting qubit quantum system, similar to those proposed for ion trap architectures \cite{Monroe08032013}. This mechanism to distribute quantum information optically, will improve the scalability of superconducting systems as size constraints imposed by the volume of the cryogenic system will no longer be an issue for large scale systems.

\acknowledgments{Funding was provided by The Johns Hopkins University Applied Physics Laboratory Internal Research and Development program. Many thanks to Shannon Nelson for assistance with the design of Fig. \ref{fig:Schematic}. I had many useful and stimulating discussion on this topic with Joan Hoffmann, Scott Hendrickson, and Layne Churchill}

\appendix
\section{Linearized Optomechanical Hamiltonian Derivation}\label{sec:h_derivation}
The linearized optomechanical Hamiltonian given in Eq. \eqref{eq:h_om} describes the interaction of a mechanical oscillator with an optical cavity driven by an external field.  The interaction arises from the radiation pressure force imparted on the mechanical oscillator from the light driving the optical cavity.  The total Hamiltonian $\hat{H} = \hat{H}_{om} + \hat{H}_{drive}$ is two quantum harmonic oscillators together with a external driving field where
\begin{equation}
\label{eq:h_om_basic}
\hat{H}_{om} = \hbar\omega_c(\hat{x})\hat{a}^\dagger\hat{a} + \hbar \omega_m \hat{b}_m^\dagger \hat{b}_m
\end{equation}
and
\begin{equation}
\label{eq:h_drive}
\hat{H}_{drive} = i\hbar E(\hat{a}^\dagger e^{-i \omega_L t} - \hat{a} e^{i\omega_L t}),
\end{equation}
where $\omega_c(\hat{x})$ is the cavity resonant frequency $\hat{a}$ is the cavity annihilation operator; $\omega_m$ is the mechanical oscillator resonant frequency; $\hat{b}_m$ is the mechanical oscillator's annihilation operator,; $E = \sqrt{\frac{P \gamma_c}{\hbar \omega_c}}$ is related the input power of the driving field, $P$, and cavity decay rate, $\gamma_c$; and $\omega_L$ is the frequency of the driving laser.  We explicitly note the dependence on $\hat{x}$ of the cavity resonant frequency, where $\hat{x} = x_{\textnormal{zpf}}(\hat{b}_m+\hat{b}_m^\dagger)$ is the displacement one of the cavity mirrors from equilibrium due to mechanical motion, and $x_{\textnormal{zpf}} = \sqrt{\bra{0}\hat{x}^2\ket{0}}$ is the zero point fluctuation of the mechanical oscillator.  We note that we have neglected the zero-point energy of both the mechanical and optical oscillators as they do not contribute to the overall dynamics.

Non-zero displacement can occur from radiation pressure forces, thereby creating the coupling between light and mechanical motion. To derive Eq. \eqref{eq:h_om}, we begin by expanding the cavity resonant frequency in a Taylor series about the origin
\begin{equation}
\label{eq:omega_taylor}
\omega_c(\hat{x}) = \omega_c + \hat{x}\frac{\partial \omega_c}{\partial \hat{x}} + \cdots.
\end{equation}
To first order in the motion of the mirror, Eq. \eqref{eq:h_om_basic} becomes
\begin{equation}
\label{eq:h_om_first_order}
\hat{H}_{om} = \hbar\omega_c\hat{a}^\dagger\hat{a} + \hbar \omega_m\hat{b}_m^\dagger\hat{b}_m - \hbar g_0 \hat{a}^\dagger\hat{a}(\hat{b}_m + \hat{b}_m^\dagger),
\end{equation}
where $g_0 = -\frac{\partial \omega_c}{\partial \hat{x}} x_{\textnormal{zpf}} = -\omega_c x_{\textnormal{zpf}}/L$ and the second equality comes from assuming a Fabry--P\'erot cavity of equilibrium length $L$.  

We now change into a reference frame rotating at laser frequency with the unitary operator $\hat{U} = \exp(i\omega_L \hat{a}^\dagger \hat{a} t)$ via $\hat{H}^\prime = \hat{U}\hat{H}\hat{U}^\dagger - i\hbar\hat{U}\frac{\partial \hat{U}^\dagger}{\partial t}$.  One can verify that this transformation yields
\begin{equation}
\label{eq:h_om_rot}
\hat{H}^\prime = \hat{H}_0^\prime + \hat{H}_{int}^\prime + \hat{H}_{drive}^\prime
\end{equation}
where 
\begin{subequations}
\label{eq:h_int}
\begin{align}
& \hat{H}_0^\prime = -\hbar \Delta \hat{a}^\dagger \hat{a} + \hbar \omega_m\hat{b}^\dagger \hat{b}, \\
& \hat{H}_{int}^\prime = -\hbar g_0\hat{a}^\dagger\hat{a}(\hat{b}_m + \hat{b}_m), \\
& \hat{H}_{drive}^\prime = i \hbar E (\hat{a}^\dagger - \hat{a}),
\end{align}
\end{subequations}
and $\Delta = \omega_L - \omega_c$ is the detuning of the laser field from the cavity resonance.

We now proceed to linearize Eqs. \eqref{eq:h_int} by assuming we have a strong driving field.  Under this assumption we can write the cavity mode operators as a classical term plus a quantum fluctuation term $\hat{a} = \bar{\alpha} + \delta\hat{a}$ where $\bar{\alpha} = \sqrt{n}$ is the classical amplitude with mean photon number $n$ and $\delta\hat{a}$ denotes the quantum portion.  With this substitution Eqs. \eqref{eq:h_int} become
\begin{subequations}
\label{eq:h_int_lin1}
\begin{align}
& \hat{H}_0^\prime = -\hbar \Delta (\bar{\alpha}^* + \delta \hat{a}^\dagger)(\bar{\alpha}+\delta\hat{a}) + \hbar \omega_m\hat{b}^\dagger \hat{b}, \label{eq:h0p} \\
& \hat{H}_{int}^\prime = -\hbar g_0(\bar{\alpha}^* + \delta \hat{a}^\dagger)(\bar{\alpha}+\delta\hat{a}) (\hat{b}_m + \hat{b}_m^\dagger), \label{eq:hintp} \\ 
& \hat{H}_{drive}^\prime = i \hbar E (\bar{\alpha}^* + \delta\hat{a}^\dagger - \bar{\alpha} - \delta\hat{a}). \label{eq:hdrivep}
\end{align}
\end{subequations}
Within Eq. \eqref{eq:h0p}, we drop $\hbar \Delta \bar{\alpha}^2$ since it is simply a zero-point energy shift and does not contribute to the dynamics.  Within Eq. \eqref{eq:hintp} we again drop the term proportional to $|\bar{\alpha}|^2$ since it simply adds a displacement of the origin of the mechanical oscillator, and we drop the terms proportional to $\delta\hat{a}^\dagger\delta\hat{a}$ as they are smaller by a factor of $\alpha$ compared to the term we keep.  After making these assumptions, and taking $\bar{\alpha}$ to be real without loss of generality, Eqs. \eqref{eq:h_int_lin1} become
\begin{subequations}
\label{eq:h_int_lin2}
\begin{align}
& \hat{H}_0^\prime \approx -\hbar \Delta (\bar{\alpha}\delta\hat{a} + \bar{\alpha}\delta\hat{a}^\dagger + \delta \hat{a}^\dagger \delta\hat{a}) + \hbar \omega_m\hat{b}^\dagger \hat{b}, \\
& \hat{H}_{int}^\prime \approx -\hbar g(\delta \hat{a}^\dagger + \delta\hat{a}) (\hat{b}_m + \hat{b}_m^\dagger), \\ 
& \hat{H}_{drive}^\prime \approx i \hbar E (\delta\hat{a}^\dagger - \delta\hat{a}),
\end{align}
\end{subequations}
where $g = g_0\bar{\alpha}$ is the optomechanical coupling strength that is enhanced by a factor proportional to the optical driving field amplitude.

We now once again change reference frames, this time going to a frame rotating at the mechanical frequency with the operator $\hat{U} = \exp(i\omega_m \delta\hat{a}^\dagger \delta\hat{a} t + i\omega_m \hat{b}_m^\dagger \hat{b}_m t)$.  Applying this transformation yields
\begin{subequations}
\label{eq:h_int_lin3}
\begin{align}
& \hat{H}_0^{\prime\prime} \approx -\hbar (\Delta + \omega_m) \delta \hat{a}^\dagger \delta\hat{a} -\hbar \Delta (\bar{\alpha}\delta\hat{a}e^{-i\omega_m t} + \bar{\alpha}\delta\hat{a}^\dagger e^{i\omega_m t}), \\
& \hat{H}_{int}^{\prime\prime} \approx -\hbar g(\delta \hat{a}^\dagger e^{i\omega_m t} + \delta\hat{a}e^{-i\omega_m t}) (\hat{b}_m e^{-i\omega_m t} + \hat{b}_m^\dagger e^{i\omega_m t}), \\ 
& \hat{H}_{drive}^{\prime\prime} \approx i \hbar E (\delta\hat{a}^\dagger e^{i\omega_m t} - \delta\hat{a}e^{-i\omega_m t}),
\end{align}
\end{subequations}

We now make the assumption that $\Delta \approx -\omega_m$.  This allows us to employ the rotating wave approximation (RWA) and eliminate the terms oscillating at $\pm i\omega_mt$ and higher.  This yields the resulting form of the optomechanical interaction Hamiltonian
\begin{equation}
\label{eq:h_int_lin4}
\hat{H}_{int}^{\prime\prime} \approx -\hbar g(\delta\hat{a}^\dagger\hat{b}_m + \delta\hat{a}\hat{b}_m^\dagger)
\end{equation}
that we use in our analysis.  Both $\hat{H}_0^{\prime\prime}$ and $\hat{H}_{drive}^{\prime\prime}$ can be dropped in the RWA.

\section{Numerical Solution Method}\label{sec:numerical_solution}
Here we sketch our numerical solution technique used to solve Eq. \eqref{eq:solved_Langevin_general}. The propagator that must be evaluated is given by
\begin{equation}
\label{eq:propagator}
\exp\left(-i\int_{t^\prime}^t dt^{\prime\prime} M(t^{\prime})\right)
\end{equation}
where $M(t)$ is define in Eq. \eqref{eq:m_matrix}. We numerically evaluate the time ordered exponential in the standard way as
\begin{align}
\label{eq:time_ordered}
& \mathcal{T}\left\{e^{-i\int_{t^\prime}^t dt^{\prime\prime} M(t^{\prime\prime})}\right\} \\ \nonumber
& = \lim_{N\to\infty}\left[e^{M(t_N)\Delta t} e^{M(t_{N-1})\Delta t} \cdots e^{M(t_1)\Delta t} e^{M(t_0)\Delta t} \right],
\end{align}
where $t_j = j \Delta t$ and $\Delta t = t/N$ for $j = M, \hdots, N$ with $t^\prime = M t /N$ the initial time and $M<N$. For the numerical evaluation of Eq. \eqref{eq:solved_Langevin_general} we truncate the expansion given in Eq. \eqref{eq:time_ordered} such that $\Delta t \ll 1$.

To numerically evaluate the integral contained in the noise term of Eq. \eqref{eq:solved_Langevin_general} we use the trapezoidal rule
\begin{equation}
\label{eq:trapezoidal}
\int_a^b f(x) dx \approx \frac{\Delta x}{2}\sum_{k=1}^N [f(x_{k+1}) + f(x_k)],
\end{equation}
where $\Delta x = (b-a)/N$. We choose a $\Delta x \ll 1$ to ensure good convergence of the numerical integral.
%


\begin{thebibliography}{47}%
\makeatletter
\providecommand \@ifxundefined [1]{%
 \@ifx{#1\undefined}
}%
\providecommand \@ifnum [1]{%
 \ifnum #1\expandafter \@firstoftwo
 \else \expandafter \@secondoftwo
 \fi
}%
\providecommand \@ifx [1]{%
 \ifx #1\expandafter \@firstoftwo
 \else \expandafter \@secondoftwo
 \fi
}%
\providecommand \natexlab [1]{#1}%
\providecommand \enquote  [1]{``#1''}%
\providecommand \bibnamefont  [1]{#1}%
\providecommand \bibfnamefont [1]{#1}%
\providecommand \citenamefont [1]{#1}%
\providecommand \href@noop [0]{\@secondoftwo}%
\providecommand \href [0]{\begingroup \@sanitize@url \@href}%
\providecommand \@href[1]{\@@startlink{#1}\@@href}%
\providecommand \@@href[1]{\endgroup#1\@@endlink}%
\providecommand \@sanitize@url [0]{\catcode `\\12\catcode `\$12\catcode
  `\&12\catcode `\#12\catcode `\^12\catcode `\_12\catcode `\%12\relax}%
\providecommand \@@startlink[1]{}%
\providecommand \@@endlink[0]{}%
\providecommand \url  [0]{\begingroup\@sanitize@url \@url }%
\providecommand \@url [1]{\endgroup\@href {#1}{\urlprefix }}%
\providecommand \urlprefix  [0]{URL }%
\providecommand \Eprint [0]{\href }%
\providecommand \doibase [0]{http://dx.doi.org/}%
\providecommand \selectlanguage [0]{\@gobble}%
\providecommand \bibinfo  [0]{\@secondoftwo}%
\providecommand \bibfield  [0]{\@secondoftwo}%
\providecommand \translation [1]{[#1]}%
\providecommand \BibitemOpen [0]{}%
\providecommand \bibitemStop [0]{}%
\providecommand \bibitemNoStop [0]{.\EOS\space}%
\providecommand \EOS [0]{\spacefactor3000\relax}%
\providecommand \BibitemShut  [1]{\csname bibitem#1\endcsname}%
\let\auto@bib@innerbib\@empty
\bibitem [{\citenamefont {Braginsky}\ \emph {et~al.}(1970)\citenamefont
  {Braginsky}, \citenamefont {Manukin},\ and\ \citenamefont
  {Tikhonov}}]{Braginskii1970}%
  \BibitemOpen
  \bibfield  {author} {\bibinfo {author} {\bibfnamefont {V.~B.}\ \bibnamefont
  {Braginsky}}, \bibinfo {author} {\bibfnamefont {A.~B.}\ \bibnamefont
  {Manukin}}, \ and\ \bibinfo {author} {\bibfnamefont {M.~Y.}\ \bibnamefont
  {Tikhonov}},\ }\href
  {http://www.jetp.ac.ru/cgi-bin/e/index/e/31/5/p829?a=list} {\bibfield
  {journal} {\bibinfo  {journal} {Soviet Phys. JETP}\ }\textbf {\bibinfo
  {volume} {31}},\ \bibinfo {pages} {829} (\bibinfo {year} {1970})}\BibitemShut
  {NoStop}%
\bibitem [{\citenamefont {Cleland}\ and\ \citenamefont
  {Roukes}(1996)}]{apl/69/18/10.1063/1.117548}%
  \BibitemOpen
  \bibfield  {author} {\bibinfo {author} {\bibfnamefont {A.~N.}\ \bibnamefont
  {Cleland}}\ and\ \bibinfo {author} {\bibfnamefont {M.~L.}\ \bibnamefont
  {Roukes}},\ }\href {\doibase http://dx.doi.org/10.1063/1.117548} {\bibfield
  {journal} {\bibinfo  {journal} {App. Phys. Lett.}\ }\textbf {\bibinfo
  {volume} {69}},\ \bibinfo {pages} {2653} (\bibinfo {year}
  {1996})}\BibitemShut {NoStop}%
\bibitem [{\citenamefont {O'Connell}\ \emph {et~al.}(2010)\citenamefont
  {O'Connell}, \citenamefont {Hofheinz}, \citenamefont {Ansmann}, \citenamefont
  {Bialczak}, \citenamefont {Lenander}, \citenamefont {Lucero}, \citenamefont
  {Neeley}, \citenamefont {Sank}, \citenamefont {Wang}, \citenamefont {Weides},
  \citenamefont {Wenner}, \citenamefont {Martinis},\ and\ \citenamefont
  {Cleland}}]{o2010quantum}%
  \BibitemOpen
  \bibfield  {author} {\bibinfo {author} {\bibfnamefont {A.~D.}\ \bibnamefont
  {O'Connell}}, \bibinfo {author} {\bibfnamefont {M.}~\bibnamefont {Hofheinz}},
  \bibinfo {author} {\bibfnamefont {M.}~\bibnamefont {Ansmann}}, \bibinfo
  {author} {\bibfnamefont {R.~C.}\ \bibnamefont {Bialczak}}, \bibinfo {author}
  {\bibfnamefont {M.}~\bibnamefont {Lenander}}, \bibinfo {author}
  {\bibfnamefont {E.}~\bibnamefont {Lucero}}, \bibinfo {author} {\bibfnamefont
  {M.}~\bibnamefont {Neeley}}, \bibinfo {author} {\bibfnamefont
  {D.}~\bibnamefont {Sank}}, \bibinfo {author} {\bibfnamefont {H.}~\bibnamefont
  {Wang}}, \bibinfo {author} {\bibfnamefont {M.}~\bibnamefont {Weides}},
  \bibinfo {author} {\bibfnamefont {J.}~\bibnamefont {Wenner}}, \bibinfo
  {author} {\bibfnamefont {J.~A.}\ \bibnamefont {Martinis}}, \ and\ \bibinfo
  {author} {\bibfnamefont {A.~N.}\ \bibnamefont {Cleland}},\ }\href
  {http://www.nature.com/nature/journal/v464/n7289/full/nature08967.html}
  {\bibfield  {journal} {\bibinfo  {journal} {Nature}\ }\textbf {\bibinfo
  {volume} {464}},\ \bibinfo {pages} {697} (\bibinfo {year}
  {2010})}\BibitemShut {NoStop}%
\bibitem [{\citenamefont {Kippenberg}\ and\ \citenamefont
  {Vahala}(2007)}]{Kippenberg:07}%
  \BibitemOpen
  \bibfield  {author} {\bibinfo {author} {\bibfnamefont {T.~J.}\ \bibnamefont
  {Kippenberg}}\ and\ \bibinfo {author} {\bibfnamefont {K.~J.}\ \bibnamefont
  {Vahala}},\ }\href {\doibase 10.1364/OE.15.017172} {\bibfield  {journal}
  {\bibinfo  {journal} {Opt. Express}\ }\textbf {\bibinfo {volume} {15}},\
  \bibinfo {pages} {17172} (\bibinfo {year} {2007})}\BibitemShut {NoStop}%
\bibitem [{\citenamefont {Kippenberg}\ and\ \citenamefont
  {Vahala}(2008)}]{Kippenberg29082008}%
  \BibitemOpen
  \bibfield  {author} {\bibinfo {author} {\bibfnamefont {T.~J.}\ \bibnamefont
  {Kippenberg}}\ and\ \bibinfo {author} {\bibfnamefont {K.~J.}\ \bibnamefont
  {Vahala}},\ }\href {\doibase 10.1126/science.1156032} {\bibfield  {journal}
  {\bibinfo  {journal} {Science}\ }\textbf {\bibinfo {volume} {321}},\ \bibinfo
  {pages} {1172} (\bibinfo {year} {2008})}\BibitemShut {NoStop}%
\bibitem [{\citenamefont {Marquardt}\ and\ \citenamefont
  {Girvin}(2009)}]{Physics.2.40}%
  \BibitemOpen
  \bibfield  {author} {\bibinfo {author} {\bibfnamefont {F.}~\bibnamefont
  {Marquardt}}\ and\ \bibinfo {author} {\bibfnamefont {S.~M.}\ \bibnamefont
  {Girvin}},\ }\href {\doibase 10.1103/Physics.2.40} {\bibfield  {journal}
  {\bibinfo  {journal} {Physics}\ }\textbf {\bibinfo {volume} {2}},\ \bibinfo
  {pages} {40} (\bibinfo {year} {2009})}\BibitemShut {NoStop}%
\bibitem [{\citenamefont {Aspelmeyer}\ \emph {et~al.}(2010)\citenamefont
  {Aspelmeyer}, \citenamefont {Gr\"{o}blacher}, \citenamefont {Hammerer},\ and\
  \citenamefont {Kiesel}}]{Aspelmeyer:10}%
  \BibitemOpen
  \bibfield  {author} {\bibinfo {author} {\bibfnamefont {M.}~\bibnamefont
  {Aspelmeyer}}, \bibinfo {author} {\bibfnamefont {S.}~\bibnamefont
  {Gr\"{o}blacher}}, \bibinfo {author} {\bibfnamefont {K.}~\bibnamefont
  {Hammerer}}, \ and\ \bibinfo {author} {\bibfnamefont {N.}~\bibnamefont
  {Kiesel}},\ }\href {\doibase 10.1364/JOSAB.27.00A189} {\bibfield  {journal}
  {\bibinfo  {journal} {J. Opt. Soc. Am. B}\ }\textbf {\bibinfo {volume}
  {27}},\ \bibinfo {pages} {A189} (\bibinfo {year} {2010})}\BibitemShut
  {NoStop}%
\bibitem [{\citenamefont {Aspelmeyer}\ \emph {et~al.}(2013)\citenamefont
  {Aspelmeyer}, \citenamefont {Kippenberg},\ and\ \citenamefont
  {Marquardt}}]{1303.0733}%
  \BibitemOpen
  \bibfield  {author} {\bibinfo {author} {\bibfnamefont {M.}~\bibnamefont
  {Aspelmeyer}}, \bibinfo {author} {\bibfnamefont {T.~J.}\ \bibnamefont
  {Kippenberg}}, \ and\ \bibinfo {author} {\bibfnamefont {F.}~\bibnamefont
  {Marquardt}},\ }\href {http://arxiv.org/abs/1303.0733} {\enquote {\bibinfo
  {title} {Cavity optomechanics},}\ } (\bibinfo {year} {2013}),\ \Eprint
  {http://arxiv.org/abs/1303.0733} {arXiv:1303.0733} \BibitemShut {NoStop}%
\bibitem [{\citenamefont {Marquardt}\ \emph {et~al.}(2007)\citenamefont
  {Marquardt}, \citenamefont {Chen}, \citenamefont {Clerk},\ and\ \citenamefont
  {Girvin}}]{PhysRevLett.99.093902}%
  \BibitemOpen
  \bibfield  {author} {\bibinfo {author} {\bibfnamefont {F.}~\bibnamefont
  {Marquardt}}, \bibinfo {author} {\bibfnamefont {J.~P.}\ \bibnamefont {Chen}},
  \bibinfo {author} {\bibfnamefont {A.~A.}\ \bibnamefont {Clerk}}, \ and\
  \bibinfo {author} {\bibfnamefont {S.~M.}\ \bibnamefont {Girvin}},\ }\href
  {\doibase 10.1103/PhysRevLett.99.093902} {\bibfield  {journal} {\bibinfo
  {journal} {Phys. Rev. Lett.}\ }\textbf {\bibinfo {volume} {99}},\ \bibinfo
  {pages} {093902} (\bibinfo {year} {2007})}\BibitemShut {NoStop}%
\bibitem [{\citenamefont {Schliesser}\ \emph {et~al.}(2008)\citenamefont
  {Schliesser}, \citenamefont {Rivi{\`e}re}, \citenamefont {Anetsberger},
  \citenamefont {Arcizet},\ and\ \citenamefont
  {Kippenberg}}]{schliesser2008resolved}%
  \BibitemOpen
  \bibfield  {author} {\bibinfo {author} {\bibfnamefont {A.}~\bibnamefont
  {Schliesser}}, \bibinfo {author} {\bibfnamefont {R.}~\bibnamefont
  {Rivi{\`e}re}}, \bibinfo {author} {\bibfnamefont {G.}~\bibnamefont
  {Anetsberger}}, \bibinfo {author} {\bibfnamefont {O.}~\bibnamefont
  {Arcizet}}, \ and\ \bibinfo {author} {\bibfnamefont {T.~J.}\ \bibnamefont
  {Kippenberg}},\ }\href
  {http://www.nature.com/nphys/journal/v4/n5/full/nphys939.html} {\bibfield
  {journal} {\bibinfo  {journal} {Nat. Phys.}\ }\textbf {\bibinfo {volume}
  {4}},\ \bibinfo {pages} {415} (\bibinfo {year} {2008})}\BibitemShut {NoStop}%
\bibitem [{\citenamefont {Zhang}\ \emph {et~al.}(2003)\citenamefont {Zhang},
  \citenamefont {Peng},\ and\ \citenamefont {Braunstein}}]{PhysRevA.68.013808}%
  \BibitemOpen
  \bibfield  {author} {\bibinfo {author} {\bibfnamefont {J.}~\bibnamefont
  {Zhang}}, \bibinfo {author} {\bibfnamefont {K.}~\bibnamefont {Peng}}, \ and\
  \bibinfo {author} {\bibfnamefont {S.~L.}\ \bibnamefont {Braunstein}},\ }\href
  {\doibase 10.1103/PhysRevA.68.013808} {\bibfield  {journal} {\bibinfo
  {journal} {Phys. Rev. A}\ }\textbf {\bibinfo {volume} {68}},\ \bibinfo
  {pages} {013808} (\bibinfo {year} {2003})}\BibitemShut {NoStop}%
\bibitem [{\citenamefont {Verhagen}\ \emph {et~al.}(2012)\citenamefont
  {Verhagen}, \citenamefont {Del{\'e}glise}, \citenamefont {Weis},
  \citenamefont {Schliesser},\ and\ \citenamefont
  {Kippenberg}}]{verhagen2012quantum}%
  \BibitemOpen
  \bibfield  {author} {\bibinfo {author} {\bibfnamefont {E.}~\bibnamefont
  {Verhagen}}, \bibinfo {author} {\bibfnamefont {S.}~\bibnamefont
  {Del{\'e}glise}}, \bibinfo {author} {\bibfnamefont {S.}~\bibnamefont {Weis}},
  \bibinfo {author} {\bibfnamefont {A.}~\bibnamefont {Schliesser}}, \ and\
  \bibinfo {author} {\bibfnamefont {T.~J.}\ \bibnamefont {Kippenberg}},\ }\href
  {\doibase doi:10.1038/nature10787} {\bibfield  {journal} {\bibinfo  {journal}
  {Nature}\ }\textbf {\bibinfo {volume} {482}},\ \bibinfo {pages} {63}
  (\bibinfo {year} {2012})}\BibitemShut {NoStop}%
\bibitem [{\citenamefont {Gr{\"o}blacher}\ \emph {et~al.}(2009)\citenamefont
  {Gr{\"o}blacher}, \citenamefont {Hammerer}, \citenamefont {Vanner},\ and\
  \citenamefont {Aspelmeyer}}]{groblacher2009observation}%
  \BibitemOpen
  \bibfield  {author} {\bibinfo {author} {\bibfnamefont {S.}~\bibnamefont
  {Gr{\"o}blacher}}, \bibinfo {author} {\bibfnamefont {K.}~\bibnamefont
  {Hammerer}}, \bibinfo {author} {\bibfnamefont {M.~R.}\ \bibnamefont
  {Vanner}}, \ and\ \bibinfo {author} {\bibfnamefont {M.}~\bibnamefont
  {Aspelmeyer}},\ }\href {\doibase doi:10.1038/nature08171} {\bibfield
  {journal} {\bibinfo  {journal} {Nature}\ }\textbf {\bibinfo {volume} {460}},\
  \bibinfo {pages} {724} (\bibinfo {year} {2009})}\BibitemShut {NoStop}%
\bibitem [{\citenamefont {Rivi\`ere}\ \emph {et~al.}(2011)\citenamefont
  {Rivi\`ere}, \citenamefont {Del\'eglise}, \citenamefont {Weis}, \citenamefont
  {Gavartin}, \citenamefont {Arcizet}, \citenamefont {Schliesser},\ and\
  \citenamefont {Kippenberg}}]{PhysRevA.83.063835}%
  \BibitemOpen
  \bibfield  {author} {\bibinfo {author} {\bibfnamefont {R.}~\bibnamefont
  {Rivi\`ere}}, \bibinfo {author} {\bibfnamefont {S.}~\bibnamefont
  {Del\'eglise}}, \bibinfo {author} {\bibfnamefont {S.}~\bibnamefont {Weis}},
  \bibinfo {author} {\bibfnamefont {E.}~\bibnamefont {Gavartin}}, \bibinfo
  {author} {\bibfnamefont {O.}~\bibnamefont {Arcizet}}, \bibinfo {author}
  {\bibfnamefont {A.}~\bibnamefont {Schliesser}}, \ and\ \bibinfo {author}
  {\bibfnamefont {T.~J.}\ \bibnamefont {Kippenberg}},\ }\href {\doibase
  10.1103/PhysRevA.83.063835} {\bibfield  {journal} {\bibinfo  {journal} {Phys.
  Rev. A}\ }\textbf {\bibinfo {volume} {83}},\ \bibinfo {pages} {063835}
  (\bibinfo {year} {2011})}\BibitemShut {NoStop}%
\bibitem [{\citenamefont {Chan}\ \emph {et~al.}(2011)\citenamefont {Chan},
  \citenamefont {Alegre}, \citenamefont {Safavi-Naeini}, \citenamefont {Hill},
  \citenamefont {Krause}, \citenamefont {Gr{\"o}blacher}, \citenamefont
  {Aspelmeyer},\ and\ \citenamefont {Painter}}]{chan2011laser}%
  \BibitemOpen
  \bibfield  {author} {\bibinfo {author} {\bibfnamefont {J.}~\bibnamefont
  {Chan}}, \bibinfo {author} {\bibfnamefont {T.~M.}\ \bibnamefont {Alegre}},
  \bibinfo {author} {\bibfnamefont {A.~H.}\ \bibnamefont {Safavi-Naeini}},
  \bibinfo {author} {\bibfnamefont {J.~T.}\ \bibnamefont {Hill}}, \bibinfo
  {author} {\bibfnamefont {A.}~\bibnamefont {Krause}}, \bibinfo {author}
  {\bibfnamefont {S.}~\bibnamefont {Gr{\"o}blacher}}, \bibinfo {author}
  {\bibfnamefont {M.}~\bibnamefont {Aspelmeyer}}, \ and\ \bibinfo {author}
  {\bibfnamefont {O.}~\bibnamefont {Painter}},\ }\href {\doibase
  doi:10.1038/nature10461} {\bibfield  {journal} {\bibinfo  {journal} {Nature}\
  }\textbf {\bibinfo {volume} {478}},\ \bibinfo {pages} {89} (\bibinfo {year}
  {2011})}\BibitemShut {NoStop}%
\bibitem [{\citenamefont {Teufel}\ \emph {et~al.}(2008)\citenamefont {Teufel},
  \citenamefont {Harlow}, \citenamefont {Regal},\ and\ \citenamefont
  {Lehnert}}]{PhysRevLett.101.197203}%
  \BibitemOpen
  \bibfield  {author} {\bibinfo {author} {\bibfnamefont {J.~D.}\ \bibnamefont
  {Teufel}}, \bibinfo {author} {\bibfnamefont {J.~W.}\ \bibnamefont {Harlow}},
  \bibinfo {author} {\bibfnamefont {C.~A.}\ \bibnamefont {Regal}}, \ and\
  \bibinfo {author} {\bibfnamefont {K.~W.}\ \bibnamefont {Lehnert}},\ }\href
  {\doibase 10.1103/PhysRevLett.101.197203} {\bibfield  {journal} {\bibinfo
  {journal} {Phys. Rev. Lett.}\ }\textbf {\bibinfo {volume} {101}},\ \bibinfo
  {pages} {197203} (\bibinfo {year} {2008})}\BibitemShut {NoStop}%
\bibitem [{\citenamefont {Teufel}\ \emph {et~al.}(2011)\citenamefont {Teufel},
  \citenamefont {Donner}, \citenamefont {Li}, \citenamefont {Harlow},
  \citenamefont {Allman}, \citenamefont {Cicak}, \citenamefont {Sirois},
  \citenamefont {Whittaker}, \citenamefont {Lehnert},\ and\ \citenamefont
  {Simmonds}}]{teufel2011sideband}%
  \BibitemOpen
  \bibfield  {author} {\bibinfo {author} {\bibfnamefont {J.}~\bibnamefont
  {Teufel}}, \bibinfo {author} {\bibfnamefont {T.}~\bibnamefont {Donner}},
  \bibinfo {author} {\bibfnamefont {D.}~\bibnamefont {Li}}, \bibinfo {author}
  {\bibfnamefont {J.}~\bibnamefont {Harlow}}, \bibinfo {author} {\bibfnamefont
  {M.}~\bibnamefont {Allman}}, \bibinfo {author} {\bibfnamefont
  {K.}~\bibnamefont {Cicak}}, \bibinfo {author} {\bibfnamefont
  {A.}~\bibnamefont {Sirois}}, \bibinfo {author} {\bibfnamefont
  {J.}~\bibnamefont {Whittaker}}, \bibinfo {author} {\bibfnamefont
  {K.}~\bibnamefont {Lehnert}}, \ and\ \bibinfo {author} {\bibfnamefont
  {R.}~\bibnamefont {Simmonds}},\ }\href
  {http://www.nature.com/nature/journal/v475/n7356/full/nature10261.html}
  {\bibfield  {journal} {\bibinfo  {journal} {Nature}\ }\textbf {\bibinfo
  {volume} {475}},\ \bibinfo {pages} {359} (\bibinfo {year}
  {2011})}\BibitemShut {NoStop}%
\bibitem [{\citenamefont {Tian}\ and\ \citenamefont
  {Wang}(2010)}]{PhysRevA.82.053806}%
  \BibitemOpen
  \bibfield  {author} {\bibinfo {author} {\bibfnamefont {L.}~\bibnamefont
  {Tian}}\ and\ \bibinfo {author} {\bibfnamefont {H.}~\bibnamefont {Wang}},\
  }\href {\doibase 10.1103/PhysRevA.82.053806} {\bibfield  {journal} {\bibinfo
  {journal} {Phys. Rev. A}\ }\textbf {\bibinfo {volume} {82}},\ \bibinfo
  {pages} {053806} (\bibinfo {year} {2010})}\BibitemShut {NoStop}%
\bibitem [{\citenamefont {Wang}\ and\ \citenamefont
  {Clerk}(2012)}]{PhysRevLett.108.153603}%
  \BibitemOpen
  \bibfield  {author} {\bibinfo {author} {\bibfnamefont {Y.-D.}\ \bibnamefont
  {Wang}}\ and\ \bibinfo {author} {\bibfnamefont {A.~A.}\ \bibnamefont
  {Clerk}},\ }\href {\doibase 10.1103/PhysRevLett.108.153603} {\bibfield
  {journal} {\bibinfo  {journal} {Phys. Rev. Lett.}\ }\textbf {\bibinfo
  {volume} {108}},\ \bibinfo {pages} {153603} (\bibinfo {year}
  {2012})}\BibitemShut {NoStop}%
\bibitem [{\citenamefont {Tian}(2012)}]{PhysRevLett.108.153604}%
  \BibitemOpen
  \bibfield  {author} {\bibinfo {author} {\bibfnamefont {L.}~\bibnamefont
  {Tian}},\ }\href {\doibase 10.1103/PhysRevLett.108.153604} {\bibfield
  {journal} {\bibinfo  {journal} {Phys. Rev. Lett.}\ }\textbf {\bibinfo
  {volume} {108}},\ \bibinfo {pages} {153604} (\bibinfo {year}
  {2012})}\BibitemShut {NoStop}%
\bibitem [{\citenamefont {Dong}\ \emph {et~al.}(2012)\citenamefont {Dong},
  \citenamefont {Fiore}, \citenamefont {Kuzyk},\ and\ \citenamefont
  {Wang}}]{Dong21122012}%
  \BibitemOpen
  \bibfield  {author} {\bibinfo {author} {\bibfnamefont {C.}~\bibnamefont
  {Dong}}, \bibinfo {author} {\bibfnamefont {V.}~\bibnamefont {Fiore}},
  \bibinfo {author} {\bibfnamefont {M.~C.}\ \bibnamefont {Kuzyk}}, \ and\
  \bibinfo {author} {\bibfnamefont {H.}~\bibnamefont {Wang}},\ }\href {\doibase
  10.1126/science.1228370} {\bibfield  {journal} {\bibinfo  {journal}
  {Science}\ }\textbf {\bibinfo {volume} {338}},\ \bibinfo {pages} {1609}
  (\bibinfo {year} {2012})}\BibitemShut {NoStop}%
\bibitem [{\citenamefont {Barzanjeh}\ \emph {et~al.}(2011)\citenamefont
  {Barzanjeh}, \citenamefont {Vitali}, \citenamefont {Tombesi},\ and\
  \citenamefont {Milburn}}]{PhysRevA.84.042342}%
  \BibitemOpen
  \bibfield  {author} {\bibinfo {author} {\bibfnamefont {S.}~\bibnamefont
  {Barzanjeh}}, \bibinfo {author} {\bibfnamefont {D.}~\bibnamefont {Vitali}},
  \bibinfo {author} {\bibfnamefont {P.}~\bibnamefont {Tombesi}}, \ and\
  \bibinfo {author} {\bibfnamefont {G.~J.}\ \bibnamefont {Milburn}},\ }\href
  {\doibase 10.1103/PhysRevA.84.042342} {\bibfield  {journal} {\bibinfo
  {journal} {Phys. Rev. A}\ }\textbf {\bibinfo {volume} {84}},\ \bibinfo
  {pages} {042342} (\bibinfo {year} {2011})}\BibitemShut {NoStop}%
\bibitem [{\citenamefont {Tian}(2013)}]{PhysRevLett.110.233602}%
  \BibitemOpen
  \bibfield  {author} {\bibinfo {author} {\bibfnamefont {L.}~\bibnamefont
  {Tian}},\ }\href {\doibase 10.1103/PhysRevLett.110.233602} {\bibfield
  {journal} {\bibinfo  {journal} {Phys. Rev. Lett.}\ }\textbf {\bibinfo
  {volume} {110}},\ \bibinfo {pages} {233602} (\bibinfo {year}
  {2013})}\BibitemShut {NoStop}%
\bibitem [{\citenamefont {Barzanjeh}\ \emph {et~al.}(2012)\citenamefont
  {Barzanjeh}, \citenamefont {Abdi}, \citenamefont {Milburn}, \citenamefont
  {Tombesi},\ and\ \citenamefont {Vitali}}]{PhysRevLett.109.130503}%
  \BibitemOpen
  \bibfield  {author} {\bibinfo {author} {\bibfnamefont {S.}~\bibnamefont
  {Barzanjeh}}, \bibinfo {author} {\bibfnamefont {M.}~\bibnamefont {Abdi}},
  \bibinfo {author} {\bibfnamefont {G.~J.}\ \bibnamefont {Milburn}}, \bibinfo
  {author} {\bibfnamefont {P.}~\bibnamefont {Tombesi}}, \ and\ \bibinfo
  {author} {\bibfnamefont {D.}~\bibnamefont {Vitali}},\ }\href {\doibase
  10.1103/PhysRevLett.109.130503} {\bibfield  {journal} {\bibinfo  {journal}
  {Phys. Rev. Lett.}\ }\textbf {\bibinfo {volume} {109}},\ \bibinfo {pages}
  {130503} (\bibinfo {year} {2012})}\BibitemShut {NoStop}%
\bibitem [{\citenamefont {Hill}\ \emph {et~al.}(2012)\citenamefont {Hill},
  \citenamefont {Safavi-Naeini}, \citenamefont {Chan},\ and\ \citenamefont
  {Painter}}]{hill2012coherent}%
  \BibitemOpen
  \bibfield  {author} {\bibinfo {author} {\bibfnamefont {J.~T.}\ \bibnamefont
  {Hill}}, \bibinfo {author} {\bibfnamefont {A.~H.}\ \bibnamefont
  {Safavi-Naeini}}, \bibinfo {author} {\bibfnamefont {J.}~\bibnamefont {Chan}},
  \ and\ \bibinfo {author} {\bibfnamefont {O.}~\bibnamefont {Painter}},\ }\href
  {\doibase 10.1038/ncomms2201} {\bibfield  {journal} {\bibinfo  {journal}
  {Nat. Comm.}\ }\textbf {\bibinfo {volume} {3}},\ \bibinfo {pages} {1196}
  (\bibinfo {year} {2012})}\BibitemShut {NoStop}%
\bibitem [{\citenamefont {Bochmann}\ \emph {et~al.}(2013)\citenamefont
  {Bochmann}, \citenamefont {Vainsencher}, \citenamefont {Awschalom},\ and\
  \citenamefont {Cleland}}]{bochmann2013nanomechanical}%
  \BibitemOpen
  \bibfield  {author} {\bibinfo {author} {\bibfnamefont {J.}~\bibnamefont
  {Bochmann}}, \bibinfo {author} {\bibfnamefont {A.}~\bibnamefont
  {Vainsencher}}, \bibinfo {author} {\bibfnamefont {D.~D.}\ \bibnamefont
  {Awschalom}}, \ and\ \bibinfo {author} {\bibfnamefont {A.~N.}\ \bibnamefont
  {Cleland}},\ }\href {\doibase 10.1038/nphys2748} {\bibfield  {journal}
  {\bibinfo  {journal} {Nat. Phys.}\ }\textbf {\bibinfo {volume} {9}},\
  \bibinfo {pages} {712} (\bibinfo {year} {2013})}\BibitemShut {NoStop}%
\bibitem [{\citenamefont {Bagci}\ \emph {et~al.}(2014)\citenamefont {Bagci},
  \citenamefont {Simonsen}, \citenamefont {Schmid}, \citenamefont {Villanueva},
  \citenamefont {Zeuthen}, \citenamefont {Appel}, \citenamefont {Taylor},
  \citenamefont {S{\o}rensen}, \citenamefont {Usami}, \citenamefont
  {Schliesser},\ and\ \citenamefont {Polzik}}]{bagci2013optical}%
  \BibitemOpen
  \bibfield  {author} {\bibinfo {author} {\bibfnamefont {T.}~\bibnamefont
  {Bagci}}, \bibinfo {author} {\bibfnamefont {A.}~\bibnamefont {Simonsen}},
  \bibinfo {author} {\bibfnamefont {S.}~\bibnamefont {Schmid}}, \bibinfo
  {author} {\bibfnamefont {L.}~\bibnamefont {Villanueva}}, \bibinfo {author}
  {\bibfnamefont {E.}~\bibnamefont {Zeuthen}}, \bibinfo {author} {\bibfnamefont
  {J.}~\bibnamefont {Appel}}, \bibinfo {author} {\bibfnamefont {J.~M.}\
  \bibnamefont {Taylor}}, \bibinfo {author} {\bibfnamefont {A.}~\bibnamefont
  {S{\o}rensen}}, \bibinfo {author} {\bibfnamefont {K.}~\bibnamefont {Usami}},
  \bibinfo {author} {\bibfnamefont {A.}~\bibnamefont {Schliesser}}, \ and\
  \bibinfo {author} {\bibfnamefont {E.~S.}\ \bibnamefont {Polzik}},\ }\href
  {http://www.nature.com/nature/journal/v507/n7490/full/nature13029.html}
  {\bibfield  {journal} {\bibinfo  {journal} {Nature}\ }\textbf {\bibinfo
  {volume} {507}},\ \bibinfo {pages} {81} (\bibinfo {year} {2014})}\BibitemShut
  {NoStop}%
\bibitem [{\citenamefont {Devoret}\ and\ \citenamefont
  {Schoelkopf}(2013)}]{Devoret08032013}%
  \BibitemOpen
  \bibfield  {author} {\bibinfo {author} {\bibfnamefont {M.~H.}\ \bibnamefont
  {Devoret}}\ and\ \bibinfo {author} {\bibfnamefont {R.~J.}\ \bibnamefont
  {Schoelkopf}},\ }\href {\doibase 10.1126/science.1231930} {\bibfield
  {journal} {\bibinfo  {journal} {Science}\ }\textbf {\bibinfo {volume}
  {339}},\ \bibinfo {pages} {1169} (\bibinfo {year} {2013})}\BibitemShut
  {NoStop}%
\bibitem [{\citenamefont {Wallraff}\ \emph {et~al.}(2004)\citenamefont
  {Wallraff}, \citenamefont {Schuster}, \citenamefont {Blais}, \citenamefont
  {Frunzio}, \citenamefont {Huang}, \citenamefont {Majer}, \citenamefont
  {Kumar}, \citenamefont {Girvin},\ and\ \citenamefont
  {Schoelkopf}}]{wallraff2004strong}%
  \BibitemOpen
  \bibfield  {author} {\bibinfo {author} {\bibfnamefont {A.}~\bibnamefont
  {Wallraff}}, \bibinfo {author} {\bibfnamefont {D.~I.}\ \bibnamefont
  {Schuster}}, \bibinfo {author} {\bibfnamefont {A.}~\bibnamefont {Blais}},
  \bibinfo {author} {\bibfnamefont {L.}~\bibnamefont {Frunzio}}, \bibinfo
  {author} {\bibfnamefont {R.-S.}\ \bibnamefont {Huang}}, \bibinfo {author}
  {\bibfnamefont {J.}~\bibnamefont {Majer}}, \bibinfo {author} {\bibfnamefont
  {S.}~\bibnamefont {Kumar}}, \bibinfo {author} {\bibfnamefont {S.~M.}\
  \bibnamefont {Girvin}}, \ and\ \bibinfo {author} {\bibfnamefont {R.~J.}\
  \bibnamefont {Schoelkopf}},\ }\href {\doibase doi:10.1038/nature02851}
  {\bibfield  {journal} {\bibinfo  {journal} {Nature}\ }\textbf {\bibinfo
  {volume} {431}},\ \bibinfo {pages} {162} (\bibinfo {year}
  {2004})}\BibitemShut {NoStop}%
\bibitem [{\citenamefont {Pellizzari}(1997)}]{PhysRevLett.79.5242}%
  \BibitemOpen
  \bibfield  {author} {\bibinfo {author} {\bibfnamefont {T.}~\bibnamefont
  {Pellizzari}},\ }\href {\doibase 10.1103/PhysRevLett.79.5242} {\bibfield
  {journal} {\bibinfo  {journal} {Phys. Rev. Lett.}\ }\textbf {\bibinfo
  {volume} {79}},\ \bibinfo {pages} {5242} (\bibinfo {year}
  {1997})}\BibitemShut {NoStop}%
\bibitem [{\citenamefont {van Enk}\ \emph {et~al.}(1999)\citenamefont {van
  Enk}, \citenamefont {Kimble}, \citenamefont {Cirac},\ and\ \citenamefont
  {Zoller}}]{PhysRevA.59.2659}%
  \BibitemOpen
  \bibfield  {author} {\bibinfo {author} {\bibfnamefont {S.~J.}\ \bibnamefont
  {van Enk}}, \bibinfo {author} {\bibfnamefont {H.~J.}\ \bibnamefont {Kimble}},
  \bibinfo {author} {\bibfnamefont {J.~I.}\ \bibnamefont {Cirac}}, \ and\
  \bibinfo {author} {\bibfnamefont {P.}~\bibnamefont {Zoller}},\ }\href
  {\doibase 10.1103/PhysRevA.59.2659} {\bibfield  {journal} {\bibinfo
  {journal} {Phys. Rev. A}\ }\textbf {\bibinfo {volume} {59}},\ \bibinfo
  {pages} {2659} (\bibinfo {year} {1999})}\BibitemShut {NoStop}%
\bibitem [{\citenamefont {Serafini}\ \emph {et~al.}(2006)\citenamefont
  {Serafini}, \citenamefont {Mancini},\ and\ \citenamefont
  {Bose}}]{PhysRevLett.96.010503}%
  \BibitemOpen
  \bibfield  {author} {\bibinfo {author} {\bibfnamefont {A.}~\bibnamefont
  {Serafini}}, \bibinfo {author} {\bibfnamefont {S.}~\bibnamefont {Mancini}}, \
  and\ \bibinfo {author} {\bibfnamefont {S.}~\bibnamefont {Bose}},\ }\href
  {\doibase 10.1103/PhysRevLett.96.010503} {\bibfield  {journal} {\bibinfo
  {journal} {Phys. Rev. Lett.}\ }\textbf {\bibinfo {volume} {96}},\ \bibinfo
  {pages} {010503} (\bibinfo {year} {2006})}\BibitemShut {NoStop}%
\bibitem [{\citenamefont {Monroe}\ and\ \citenamefont
  {Kim}(2013)}]{Monroe08032013}%
  \BibitemOpen
  \bibfield  {author} {\bibinfo {author} {\bibfnamefont {C.}~\bibnamefont
  {Monroe}}\ and\ \bibinfo {author} {\bibfnamefont {J.}~\bibnamefont {Kim}},\
  }\href {\doibase 10.1126/science.1231298} {\bibfield  {journal} {\bibinfo
  {journal} {Science}\ }\textbf {\bibinfo {volume} {339}},\ \bibinfo {pages}
  {1164} (\bibinfo {year} {2013})}\BibitemShut {NoStop}%
\bibitem [{\citenamefont {Stannigel}\ \emph {et~al.}(2010)\citenamefont
  {Stannigel}, \citenamefont {Rabl}, \citenamefont {S\o{}rensen}, \citenamefont
  {Zoller},\ and\ \citenamefont {Lukin}}]{PhysRevLett.105.220501}%
  \BibitemOpen
  \bibfield  {author} {\bibinfo {author} {\bibfnamefont {K.}~\bibnamefont
  {Stannigel}}, \bibinfo {author} {\bibfnamefont {P.}~\bibnamefont {Rabl}},
  \bibinfo {author} {\bibfnamefont {A.~S.}\ \bibnamefont {S\o{}rensen}},
  \bibinfo {author} {\bibfnamefont {P.}~\bibnamefont {Zoller}}, \ and\ \bibinfo
  {author} {\bibfnamefont {M.~D.}\ \bibnamefont {Lukin}},\ }\href {\doibase
  10.1103/PhysRevLett.105.220501} {\bibfield  {journal} {\bibinfo  {journal}
  {Phys. Rev. Lett.}\ }\textbf {\bibinfo {volume} {105}},\ \bibinfo {pages}
  {220501} (\bibinfo {year} {2010})}\BibitemShut {NoStop}%
\bibitem [{\citenamefont {Stannigel}\ \emph {et~al.}(2011)\citenamefont
  {Stannigel}, \citenamefont {Rabl}, \citenamefont {S\o{}rensen}, \citenamefont
  {Lukin},\ and\ \citenamefont {Zoller}}]{PhysRevA.84.042341}%
  \BibitemOpen
  \bibfield  {author} {\bibinfo {author} {\bibfnamefont {K.}~\bibnamefont
  {Stannigel}}, \bibinfo {author} {\bibfnamefont {P.}~\bibnamefont {Rabl}},
  \bibinfo {author} {\bibfnamefont {A.~S.}\ \bibnamefont {S\o{}rensen}},
  \bibinfo {author} {\bibfnamefont {M.~D.}\ \bibnamefont {Lukin}}, \ and\
  \bibinfo {author} {\bibfnamefont {P.}~\bibnamefont {Zoller}},\ }\href
  {\doibase 10.1103/PhysRevA.84.042341} {\bibfield  {journal} {\bibinfo
  {journal} {Phys. Rev. A}\ }\textbf {\bibinfo {volume} {84}},\ \bibinfo
  {pages} {042341} (\bibinfo {year} {2011})}\BibitemShut {NoStop}%
\bibitem [{\citenamefont {Nakajima}\ \emph {et~al.}(1994)\citenamefont
  {Nakajima}, \citenamefont {Elk}, \citenamefont {Zhang},\ and\ \citenamefont
  {Lambropoulos}}]{PhysRevA.50.R913}%
  \BibitemOpen
  \bibfield  {author} {\bibinfo {author} {\bibfnamefont {T.}~\bibnamefont
  {Nakajima}}, \bibinfo {author} {\bibfnamefont {M.}~\bibnamefont {Elk}},
  \bibinfo {author} {\bibfnamefont {J.}~\bibnamefont {Zhang}}, \ and\ \bibinfo
  {author} {\bibfnamefont {P.}~\bibnamefont {Lambropoulos}},\ }\href {\doibase
  10.1103/PhysRevA.50.R913} {\bibfield  {journal} {\bibinfo  {journal} {Phys.
  Rev. A}\ }\textbf {\bibinfo {volume} {50}},\ \bibinfo {pages} {R913}
  (\bibinfo {year} {1994})}\BibitemShut {NoStop}%
\bibitem [{\citenamefont {Carroll}\ and\ \citenamefont
  {Hioe}(1992)}]{PhysRevLett.68.3523}%
  \BibitemOpen
  \bibfield  {author} {\bibinfo {author} {\bibfnamefont {C.~E.}\ \bibnamefont
  {Carroll}}\ and\ \bibinfo {author} {\bibfnamefont {F.~T.}\ \bibnamefont
  {Hioe}},\ }\href {\doibase 10.1103/PhysRevLett.68.3523} {\bibfield  {journal}
  {\bibinfo  {journal} {Phys. Rev. Lett.}\ }\textbf {\bibinfo {volume} {68}},\
  \bibinfo {pages} {3523} (\bibinfo {year} {1992})}\BibitemShut {NoStop}%
\bibitem [{\citenamefont {Carroll}\ and\ \citenamefont
  {Hioe}(1993)}]{PhysRevA.47.571}%
  \BibitemOpen
  \bibfield  {author} {\bibinfo {author} {\bibfnamefont {C.~E.}\ \bibnamefont
  {Carroll}}\ and\ \bibinfo {author} {\bibfnamefont {F.~T.}\ \bibnamefont
  {Hioe}},\ }\href {\doibase 10.1103/PhysRevA.47.571} {\bibfield  {journal}
  {\bibinfo  {journal} {Phys. Rev. A}\ }\textbf {\bibinfo {volume} {47}},\
  \bibinfo {pages} {571} (\bibinfo {year} {1993})}\BibitemShut {NoStop}%
\bibitem [{\citenamefont {Parkins}\ \emph {et~al.}(1995)\citenamefont
  {Parkins}, \citenamefont {Marte}, \citenamefont {Zoller}, \citenamefont
  {Carnal},\ and\ \citenamefont {Kimble}}]{PhysRevA.51.1578}%
  \BibitemOpen
  \bibfield  {author} {\bibinfo {author} {\bibfnamefont {A.~S.}\ \bibnamefont
  {Parkins}}, \bibinfo {author} {\bibfnamefont {P.}~\bibnamefont {Marte}},
  \bibinfo {author} {\bibfnamefont {P.}~\bibnamefont {Zoller}}, \bibinfo
  {author} {\bibfnamefont {O.}~\bibnamefont {Carnal}}, \ and\ \bibinfo {author}
  {\bibfnamefont {H.~J.}\ \bibnamefont {Kimble}},\ }\href {\doibase
  10.1103/PhysRevA.51.1578} {\bibfield  {journal} {\bibinfo  {journal} {Phys.
  Rev. A}\ }\textbf {\bibinfo {volume} {51}},\ \bibinfo {pages} {1578}
  (\bibinfo {year} {1995})}\BibitemShut {NoStop}%
\bibitem [{\citenamefont {Fleischhauer}\ \emph {et~al.}(2005)\citenamefont
  {Fleischhauer}, \citenamefont {Imamoglu},\ and\ \citenamefont
  {Marangos}}]{RevModPhys.77.633}%
  \BibitemOpen
  \bibfield  {author} {\bibinfo {author} {\bibfnamefont {M.}~\bibnamefont
  {Fleischhauer}}, \bibinfo {author} {\bibfnamefont {A.}~\bibnamefont
  {Imamoglu}}, \ and\ \bibinfo {author} {\bibfnamefont {J.~P.}\ \bibnamefont
  {Marangos}},\ }\href {\doibase 10.1103/RevModPhys.77.633} {\bibfield
  {journal} {\bibinfo  {journal} {Rev. Mod. Phys.}\ }\textbf {\bibinfo {volume}
  {77}},\ \bibinfo {pages} {633} (\bibinfo {year} {2005})}\BibitemShut
  {NoStop}%
\bibitem [{\citenamefont {Gardiner}\ and\ \citenamefont
  {Zoller}(2004)}]{gardiner2004quantum}%
  \BibitemOpen
  \bibfield  {author} {\bibinfo {author} {\bibfnamefont {C.}~\bibnamefont
  {Gardiner}}\ and\ \bibinfo {author} {\bibfnamefont {P.}~\bibnamefont
  {Zoller}},\ }\href@noop {} {\emph {\bibinfo {title} {Quantum Noise}}},\
  \bibinfo {edition} {3rd}\ ed.\ (\bibinfo  {publisher} {Springer-Verlag},\
  \bibinfo {address} {Berlin},\ \bibinfo {year} {2004})\BibitemShut {NoStop}%
\bibitem [{\citenamefont {Diddams}(2010)}]{Diddams:10}%
  \BibitemOpen
  \bibfield  {author} {\bibinfo {author} {\bibfnamefont {S.~A.}\ \bibnamefont
  {Diddams}},\ }\href {\doibase 10.1364/JOSAB.27.000B51} {\bibfield  {journal}
  {\bibinfo  {journal} {J. Opt. Soc. Am. B}\ }\textbf {\bibinfo {volume}
  {27}},\ \bibinfo {pages} {B51} (\bibinfo {year} {2010})}\BibitemShut
  {NoStop}%
\bibitem [{\citenamefont {Kippenberg}\ \emph {et~al.}(2011)\citenamefont
  {Kippenberg}, \citenamefont {Holzwarth},\ and\ \citenamefont
  {Diddams}}]{Kippenberg29042011}%
  \BibitemOpen
  \bibfield  {author} {\bibinfo {author} {\bibfnamefont {T.~J.}\ \bibnamefont
  {Kippenberg}}, \bibinfo {author} {\bibfnamefont {R.}~\bibnamefont
  {Holzwarth}}, \ and\ \bibinfo {author} {\bibfnamefont {S.~A.}\ \bibnamefont
  {Diddams}},\ }\href {\doibase 10.1126/science.1193968} {\bibfield  {journal}
  {\bibinfo  {journal} {Science}\ }\textbf {\bibinfo {volume} {332}},\ \bibinfo
  {pages} {555} (\bibinfo {year} {2011})}\BibitemShut {NoStop}%
\bibitem [{\citenamefont {Loudon}(2000)}]{loudon2000quantum}%
  \BibitemOpen
  \bibfield  {author} {\bibinfo {author} {\bibfnamefont {R.}~\bibnamefont
  {Loudon}},\ }\href@noop {} {\emph {\bibinfo {title} {The quantum theory of
  light}}},\ \bibinfo {edition} {3rd}\ ed.\ (\bibinfo  {publisher} {Oxford
  University Press},\ \bibinfo {address} {New York},\ \bibinfo {year} {2000})\
  Chap.~\bibinfo {chapter} {5}\BibitemShut {NoStop}%
\bibitem [{\citenamefont {Jozsa}(1994)}]{doi:10.1080/09500349414552171}%
  \BibitemOpen
  \bibfield  {author} {\bibinfo {author} {\bibfnamefont {R.}~\bibnamefont
  {Jozsa}},\ }\href {\doibase 10.1080/09500349414552171} {\bibfield  {journal}
  {\bibinfo  {journal} {J. Mod. Optic.}\ }\textbf {\bibinfo {volume} {41}},\
  \bibinfo {pages} {2315} (\bibinfo {year} {1994})}\BibitemShut {NoStop}%
\bibitem [{\citenamefont {Isar}(2009)}]{isar2009}%
  \BibitemOpen
  \bibfield  {author} {\bibinfo {author} {\bibfnamefont {A.}~\bibnamefont
  {Isar}},\ }\href {\doibase 10.1134/S1547477109070164} {\bibfield  {journal}
  {\bibinfo  {journal} {Phys. Part. Nuclei Lett.}\ }\textbf {\bibinfo {volume}
  {6}},\ \bibinfo {pages} {567} (\bibinfo {year} {2009})}\BibitemShut {NoStop}%
\bibitem [{\citenamefont {Julsgaard}\ and\ \citenamefont
  {M\o{}lmer}(2014)}]{PhysRevA.89.012333}%
  \BibitemOpen
  \bibfield  {author} {\bibinfo {author} {\bibfnamefont {B.}~\bibnamefont
  {Julsgaard}}\ and\ \bibinfo {author} {\bibfnamefont {K.}~\bibnamefont
  {M\o{}lmer}},\ }\href {\doibase 10.1103/PhysRevA.89.012333} {\bibfield
  {journal} {\bibinfo  {journal} {Phys. Rev. A}\ }\textbf {\bibinfo {volume}
  {89}},\ \bibinfo {pages} {012333} (\bibinfo {year} {2014})}\BibitemShut
  {NoStop}%
\end{thebibliography}
\end{document}